\documentclass[a4paper,11pt]{article}

\usepackage{ragged2e} 
\usepackage{changepage}

\usepackage{siunitx} 
\usepackage{amsmath}
\usepackage{bm} 
\usepackage{amssymb}
\usepackage{wasysym}
\usepackage{xfrac} 

\usepackage{graphicx}
\usepackage{subfig}
\usepackage{sidecap}
\usepackage{wrapfig}
\usepackage{caption}
\usepackage{verbatim}
\usepackage[english]{babel}

\usepackage{array}
\usepackage{adjustbox}
\usepackage{float}
\usepackage{enumerate}
\usepackage{multirow}
\usepackage{multicol}
\usepackage[table,xcdraw]{xcolor}
\usepackage[round]{natbib}

\usepackage{csvsimple}
\usepackage{bbm}
\usepackage{pdflscape}
\usepackage{breqn}
\usepackage{lscape}
\usepackage{amsfonts}
\usepackage{latexsym}
 
\usepackage[titletoc]{appendix}  
\usepackage[colorlinks=true, citecolor= blue]{hyperref} 
\hypersetup{linkcolor=black} 
\usepackage[all]{hypcap} 

\usepackage{enumitem}

\usepackage{cleveref}

\usepackage{setspace}
\usepackage{relsize} 

\usepackage{authblk}

\usepackage{tabularx}

\title{\vspace{-1cm} \huge \textbf{A Low-Rank Bayesian Approach for Geoadditive Modeling}}
\date{}

\author[1,2]{Bryan Sumalinab\footnote{Corresponding author. {\textit{E-mail address}: bryan.sumalinab@uhasselt.be}}}
\author[1]{Oswaldo Gressani}
\author[1,3]{Niel Hens}
\author[1]{Christel Faes}

\affil[1]{Interuniversity Institute for Biostatistics and statistical Bioinformatics (I-BioStat), Data Science Institute (DSI), Hasselt University, Hasselt, Belgium}
\affil[2]{Department of Mathematics and Statistics, College of Science and Mathematics, Mindanao State University - Iligan Institute of Technology, Iligan City, Philippines}
\affil[3]{Centre for Health Economic Research and Modelling Infectious Diseases (CHERMID), Vaccine \& Infectious Disease Institute, Antwerp University, Antwerp, Belgium}

\begin{document}
\maketitle

\newpage

\begin{abstract}
  \noindent Kriging is an established methodology for predicting spatial data in geostatistics. Current kriging techniques can handle linear dependencies on spatially referenced covariates. Although splines have shown promise in capturing nonlinear dependencies of covariates, their combination with kriging, especially in handling count data, remains underexplored. This paper proposes a novel Bayesian approach to the low-rank representation of geoadditive models, which integrates splines and kriging to account for both spatial correlations and nonlinear dependencies of covariates. The proposed method accommodates Gaussian and count data inherent in many geospatial datasets. Additionally, Laplace approximations to selected posterior distributions enhances computational efficiency, resulting in faster computation times compared to Markov chain Monte Carlo techniques commonly used for Bayesian inference. Method performance is assessed through a simulation study, demonstrating the effectiveness of the proposed approach. The methodology is applied to the analysis of heavy metal concentrations in the Meuse river and vulnerability to the coronavirus disease 2019 (COVID-19) in Belgium. Through this work, we provide a new flexible and computationally efficient framework for analyzing spatial data.\\

  
  \thispagestyle{empty}
\end{abstract}

\newpage
\clearpage\null

\pagenumbering{arabic}
\section{Introduction}

Observations characterized by spatial locations often exhibit inherent correlations, with closer observations demonstrating stronger dependencies than those farther apart. This spatial correlation phenomenon is a fundamental aspect of spatial data analysis, especially in disciplines such as geostatistics, where the spatial arrangement of data points provides valuable insights into underlying processes where proximity often implies similarity. \cite{cressie1993} categorized spatial data into three main types: areal (or lattice) data, geostatistical (continuous) data, and point patterns. Our primary focus here is on the analysis of geostatistical data. Geostatistics is a field dedicated to studying phenomena that are continuously distributed over spatial domains. For example, in environmental studies, nearby soil samples are likely to have similar characteristics due to shared environmental conditions and geological processes. The principles and methodologies of geostatistics can also be applied to phenomena that are not strictly continuous (e.g. areal data), including those related to epidemic modeling and public health outcomes. For example, geostatistical methods can be used to assess the impact of spatially referenced exposure/covariates on health outcomes, model the spatial variation in disease incidence rates, identify high-risk areas, and assess the spatial dependence of disease transmission \citep[see e.g.][]{waller2004, diggle2019}. In addition, spatial interpolation techniques, similar to those used in geostatistics, can also help to estimate disease prevalence in regions with sparse or missing data.\\

One of the primary applications in geostatistics is spatial prediction/interpolation, where missing or unobserved values at specific (unsampled) locations within a spatial domain are estimated. Kriging, a widely used geostatistical technique, relies on spatial correlations to interpolate and predict values across a spatial domain. Kriging methods estimate the spatial variability by considering the spatial arrangement and correlation between observations, resulting in predictions that minimize estimation errors. The strength of kriging lies in its ability to incorporate both the spatial trend and the spatial correlation structure of the data, making it a powerful tool for spatial data analysis. Although theories about kriging are well established, dealing with larger sample sizes presents a significant computational burden, which is particularly evident in the inversion of the covariance matrix. The computational cost increases as the dimension of the covariance matrix grows with sample size. A promising solution comes in the form of low-rank representations of spatial models using basis functions, which substantially improve computation time. These low-rank approaches, reviewed by \cite{wikle2010lowrank} and \cite{cressie2022}, offer a practical means to handle large datasets. One such low-rank kriging approach is proposed by \cite{kammann2003}, and involves reducing the spatial locations with a subset, called ``knots'', using a space-filling algorithm \citep{johnson1990, nychka1998}.  They adopt a spline-basis approach and rely upon the commonly used stationary covariance matrix in kriging as the basis functions. Their method offers not only computational advantages for handling big data but also ease of implementation through standard mixed models software. In addition to kriging, other techniques and modeling approaches are employed to address specific challenges and objectives. Generalized additive models \citep{wood2017}, for instance, offer flexible frameworks for capturing nonlinear relationships in the data. This approach was also implemented by \cite{kammann2003} in combination with spatial smoothing, which they termed as geoadditive modeling. \cite{vandendijck2017} extended their method by proposing to estimate the spatial decay parameter. Both approaches \citep{kammann2003,vandendijck2017} use likelihood-based estimation methods through mixed model representations of splines \citep{wand2003, ruppert2003} and are implemented in the context of Gaussian data.\\

This paper proposes a Bayesian approach for geoadditive modeling where spatial components are modeled in a similar way as in \cite{kammann2003}. The Laplace approximation is used to approximate the posterior distribution of regression parameters, so as to significantly reduce the computational time to carry out inference as compared to traditional Markov chain Monte Carlo (MCMC) algorithms. Penalized B-splines (P-splines) \citep{eilers1996} are used to model the nonlinear effects of covariates. This smoother offers the advantage of a penalty matrix that can be easily constructed and naturally translated into a Bayesian framework \citep{lang2004}. The combination of Laplace approximations and P-splines (Laplacian-P-splines) in generalized additive models developed by \cite{gressani2021} offers a computationally efficient alternative to classic MCMC approaches and serve as a backbone to the methodology developed here. We extend the Laplacian-P-splines methodology to a geoadditive model. Additionally, a novel approach is proposed for handling count data in combination with linear and nonlinear dependencies on covariates, an aspect not well explored in the literature on geostatistical modeling. Typically, a Poisson distribution is assumed for count data. However, this assumption is inadequate for handling overdispersion, where the variability exceeds the mean, as the Poisson model requires the mean and variance to be equal. In contrast, the negative binomial distribution, although more complex, accounts for overdispersion and permits more sophisticated modeling of count data. This paper implements both Poisson and negative binomial distributions, providing a robust and flexible approach for handling spatial count data.\\

The article is organized as follows. Section 2 introduces the geoadditive model, explaining how the smooth covariates and spatial components are modeled. It also presents the Bayesian geoadditive model and discusses the use of Laplace approximations, optimization of hyperparameters, predictions, criteria for model selection, and hypothesis testing for the significance of nonlinear covariates. Section 3 conducts a simulation study to assess the proposed methodology using various performance measures, including bias, relative bias, and credible and prediction interval coverage. In Section 4, the proposed model is applied to the analysis of two real-world datasets: the Meuse river data and the coronavirus disease 2019 (COVID-19) vulnerability data in Belgium. Finally, Section 5 concludes the paper. Code to reproduce results of this article is available on the following repository \href{https://github.com/bryansumalinab/Geoadditive-Modeling.git}{\color{blue}{https://github.com/bryansumalinab/Geoadditive-Modeling.git}}.

\section{Important concepts and methodology}

\subsection{Geoadditive model}
Consider spatially indexed observations denoted by $y_i(\boldsymbol{w}_i)$, where $\boldsymbol{w}_i = (w_{1i}, w_{2i}) ^{\top} \in \mathbb{R}^2$ denotes the spatial coordinates for $i = 1,...,n$. The observations $y_i(\boldsymbol{w}_i)$ are typically assumed to have a distribution from an exponential family as in generalized linear models. The geoadditive model can be written as:
\begin{equation}
\label{eqn:genmod}
    g(\mu_i) = \underbrace{\beta_0 + \beta_1x_{i1} + \cdots \beta_px_{ip}}_{\textrm{Linear predictors}} \quad  + \quad \underbrace{ f_1(s_{i1}) +  \cdots f_q(s_{iq})}_{\textrm{Smooth terms}} \quad + \underbrace{s(\boldsymbol{w}_i)}_{\textrm{Spatial component}},
\end{equation}
where $g(\cdot)$ is the link function and $\mathbb{E}(y_i(\boldsymbol{w}_i)) = \mu_i$. Model \eqref{eqn:genmod} consists of three different components. The first component contains the linear predictors $(1, x_{i1}, \dots, x_{ip})$, with corresponding parameters $(\beta_{0}, \beta_{1}, \dots, \beta_{p})$. The second component captures nonlinear dependencies of $g(\mu_i)$ on covariates $s_{ij}$, for $j=1,\dots ,q$. Each smooth covariate $s_{ij}$ can be modeled as: 
$$f_j(s_{ij})=\sum_{k=1}^K \theta_{jk} b_{jk}(s_{ij}), \quad \quad j=1,\dots,q,$$ 
where $b_{jk}(\cdot)$ is a basis function and $\theta_{jk}$ is the coefficient for $k=1,\dots,K$. In our case, the B-spline basis function is used with a discrete difference penalty on successive B-spline coefficients proposed by \cite{eilers1996}. The penalty controls the roughness of the fit and can be naturally extended to the Bayesian framework in formulating the joint prior distribution of the B-spline coefficients \citep{lang2004}. That is, if we let $\boldsymbol{\theta}_j = (\theta_{j1}, \theta_{j2}, \dots , \theta_{jK})^{\top}$, then a Gaussian prior distribution is assumed on $\boldsymbol{\theta}_j$ given by $(\boldsymbol{\theta}_j|\lambda_j) \sim \mathcal{N}_{K}(\boldsymbol{0},(\lambda_j P)^{-1})$ where $\lambda_j>0$ is the penalty parameter for the \textit{j}th smooth model component and $P = D_m^{\top}D_m$ where $D_m$ denotes the \textit{m}th order difference matrix ($m=2$ in this paper). To ensure that the penalty matrix is of full rank, a diagonal matrix is added to $P$ with small entries on the main diagonal (e.g. $10^{-12}$).\\

The third component of \eqref{eqn:genmod}, $s(\boldsymbol{w}_i)$, accounts for spatial correlation and can be modeled in several ways depending on the type of spatial data. In kriging or classical geostatistics, the observations $y_i(\boldsymbol{w}_i)$ are assumed to be continuous in the spatial domain $\boldsymbol{w}_i$, and $s(\boldsymbol{w}_i)$ is assumed to be a Gaussian process with mean 0 and variance $\sigma^{2}_{w}$. An important assumption of kriging is that $s(\boldsymbol{w}_i)$, for $i=1,\dots,n$, are correlated such that $\text{Cov}(s(\boldsymbol{w}_i), s(\boldsymbol{w}_j)) = R(\boldsymbol{w}_i - \boldsymbol{w}_j)$, which satisfies the stationarity assumption since the covariance function $R(\cdot)$ only depends on the distances between spatial locations. The main goal of kriging is to predict observations at a given location. For Gaussian data, the best linear unbiased prediction for an arbitrary location is analytically available along with the corresponding prediction uncertainty \citep{zimmerman2010classical}. For non-Gaussian data, one can rely, for example, on Bayesian \citep{wikle2010hierarchical} or likelihood-based methodologies \citep{zimmerman2010likelihood}.\\

Kriging predictions require to compute the inverse of the covariance matrix $R(\cdot)$ which is of dimension $n \times n$. With increasing sample size, the computational burden associated to these predictions becomes non-negligible. One way to address this problem is to write the spatial component in terms of the basis function model as follows:
\begin{equation}
\label{eqn:lowrankspat}
s(\cdot) = \sum_{l=1}^{L} \alpha_l b_{l}(\cdot) + \varepsilon,
\end{equation}
where $b_{l}(\cdot)$ is a known basis function and $\alpha_{l}$ are the coefficients for $l=1,\dots,L$ with $L < n$. Define $\boldsymbol{\alpha} = (\alpha_1,\dots,\alpha_L)^{\top}$. The vector $\boldsymbol{\alpha}$ is assumed to have a multivariate Gaussian distribution, that is, $\boldsymbol{\alpha} \sim \mathcal{N}_{L}(0, \Sigma_{\boldsymbol{\alpha}})$ where the form of $\Sigma_{\boldsymbol{\alpha}}$ depends on the choice of the basis function and the error term $\varepsilon$ is usually assumed to have a Gaussian distribution with mean 0 and constant variance. The addition of the error term $\varepsilon$ accounts for the errors stemming from approximating the underlying spatial process with a finite set of basis functions or its low-dimensional representation \citep{wikle2010lowrank, cressie2022}. Several choices can be made for the basis functions. One such choice is the method proposed by \cite{kammann2003} using the stationary covariance function in kriging as the basis function. By replacing the coordinates by a set of points, called knots, this approach allows for a low-rank representation of the covariance function. The spatial component $s(\boldsymbol{w}_i)$ is modeled as follows:
\begin{equation}
\label{eqn:spatmodel}
    s(\boldsymbol{w}_i) = \beta_{w_{1}}w_{1i}  + \beta_{w_{2}}w_{2i} + \sum_{s=1}^{S} z_{is}(\rho) u_s,
\end{equation}
where $w_{1i}$ and $w_{2i}$ are the spatial coordinates with corresponding coefficients  $\beta_{w_{1}}$ and $\beta_{w_{2}}$, $z_{is}(\rho) = R_{\rho}(\boldsymbol{w}_i - \boldsymbol{\kappa}_s)$ and $ \boldsymbol{u} = (u_{1}, u_{2},\dots, u_{S})^{\top}$ are assumed to be normally distributed such that ($\boldsymbol{u}|\lambda_{spat},\rho) \sim \mathcal{N}_{S}(0, (\lambda_{spat}\Omega_{\rho})^{-1})$, where $\lambda_{spat}>0$ and $\Omega_{\rho} = R_{\rho}(\boldsymbol{\kappa}_s - \boldsymbol{\kappa}_{s'})$ is an $S \times S$ matrix for all $s, s' \in {1, \dots, S}$. Here, $R_{\rho}(\cdot)$ is the covariance function used in kriging and $\boldsymbol{\kappa}_s$ ($s=1,\dots,S$) is a subset of the spatial coordinates. One way to efficiently choose these two-dimensional knots ($\boldsymbol{\kappa}_s$) is through the use of a space-filling algorithm  \citep{johnson1990, nychka1998}. In addition, there is an additional parameter, $\rho$, representing the range parameter in kriging, see e.g. \cite{fahrmeir2013} pp. 453 - 456, for commonly used covariance functions. The covariance functions used in this paper are:
\[
\begin{aligned}
    & \text{Exponential}: R_{\rho}(\boldsymbol{d}) = \lambda_{spat}^{-1} \exp(-\rho||\boldsymbol{d}||),\\
    & \text{Mat\'ern}: R_{\rho}(\boldsymbol{d}) = \lambda_{spat}^{-1} \exp(-\rho||\boldsymbol{d}||)(1+\rho||\boldsymbol{d}||),\\
     & \text{Spherical}: R_{\rho}(\boldsymbol{d}) = \lambda_{spat}^{-1} (1 - 1.5\rho||\boldsymbol{d}|| \; + \; 0.5 \rho^{3}||\boldsymbol{d}||^{3}) \; {\mathbb{I}(||\boldsymbol{d}|| \leq \rho^{-1}}),\\
      & \text{Circular}: R_{\rho}(\boldsymbol{d}) = \lambda_{spat}^{-1} \exp(-\rho^{2}||\boldsymbol{d}||^{2}),
\end{aligned}
\]
where $||\cdot||$ refers to the Euclidean distance and $\mathbb{I}(\cdot)$ is the indicator function. In the Gaussian scenario, it becomes apparent that even without the inclusion of the additional error term as in \eqref{eqn:lowrankspat}, model \eqref{eqn:spatmodel} is capable of generating accurate predictions for $y_{i}(\boldsymbol{w}_i)$. This is primarily due to the fact that any extra variance introduced by $\varepsilon$ tends to be absorbed by the measurement error variance inherent in the Gaussian model. However, the same cannot be said for count data when the Poisson distribution is assumed. To address this, we propose a negative binomial model for the count data. In this way, the error term $\varepsilon$ is regarded as excess variability, which can be effectively managed by the overdispersion parameter in the negative binomial model. This allows for a more accurate estimation of the prediction uncertainty. However, if this excess variability is small, then the Poisson assumption may be sufficient. It is also important to note that equation \eqref{eqn:spatmodel} represents a model for the average spatial process.

\subsection{Bayesian model formulation}
For ease of notation let $y_i := y_{i}(\boldsymbol{w}_i)$. To generalize the derivations, we can write the probability distributions of $y_i$ as an exponential dispersion family, $y_i \sim \text{EDF}(\gamma_i, \phi)$ with probability distribution given by $ p(y_i;\gamma, \phi) = \exp\{[y_i\gamma_i - b(\gamma_i)]/a(\phi) + c(y_i,\phi)\}$ where $\phi$ is a dispersion parameter and $\gamma_i$ is a natural parameter with mean $\mathbb{E}(y_i) = b'(\gamma_i) = \mu_i$ and variance $\mathbb{V}(y_i) = a(\phi)s''(\gamma_i)$.\\

The log-link function is used for the Poisson and negative binomial model such that $g(\mu_i)=\log(\mu_i)$. Note that an offset term $N_i$ (e.g., number of populations) may be added so that $\mu_i = \exp(g(\mu_i)) \times N_i$. In matrix form, \eqref{eqn:genmod} can be written as:

\begin{equation}
\log(\boldsymbol{\mu}) = X\boldsymbol{\beta} + \sum_{j=1}^{q}B_{j}(s_j) \boldsymbol{\theta}_j + Z(\rho)\boldsymbol{u},
    \label{eqn:matrixmod}
\end{equation}

\sloppy where $\boldsymbol{\mu} = (\mu_1, \dots, \mu_n)^{\top}$, $X$ is an $n \times (p +3)$ matrix with  \textit{i}th row vector $\boldsymbol{x}_i = (1, x_{i1}, x_{i2}, \dots, x_{ip}, w_{1i}, w_{2i})^{\top}$  and coefficient vector $\boldsymbol{\beta} = (\beta_{0}, \beta_{1}, \beta_{2},\dotsm \beta_{p}, \beta_{w_{1}}, \beta_{w_{2}})^{\top}$, $B_j$ is an $n \times K$ matrix with \textit{i}th row $\boldsymbol{b}_{ij} = (b_{j1}(s_{ij}), b_{j2}(s_{ij}), \dots, b_{jK}(s_{ij}))^{\top}$ and coefficients $\boldsymbol{\theta}_j = (\theta_{j1}, \theta_{j2}, \dots , \theta_{jK})^{\top}$, $Z(\rho)$ is an $n \times S$ matrix with \textit{i}th row $\boldsymbol{z}_{i}(\rho) = (z_{i1}(\rho), z_{i2}(\rho), \dots, z_{iS}(\rho))^{\top}$ and coefficients $\boldsymbol{u} = (u_{1}, u_{2}, \dots, u_{S})^{\top}$ for $i = 1,\dots, n$ and $j=1,\dots, q$.\\

A Gaussian prior is assumed for $\boldsymbol{\beta}$, i.e., $\boldsymbol{\beta} \sim \mathcal{N}_{\text{dim}(\boldsymbol{\beta})}(0, V_{\boldsymbol{\beta}}^{-1})$ where $V_{\boldsymbol{\beta}} = \zeta I$ with small $\zeta$ (e.g. $\zeta = 10^{-5}$ in this paper).  Denote the global design matrix as $C_{\rho} = [X:B_{1}:B_{2}:\cdots:B_{q}:Z(\rho)]$ and the corresponding parameter vector $\boldsymbol{\xi} = (\boldsymbol{\beta}^{\top}, \boldsymbol{\theta}_{1}^{\top}, \dots, \boldsymbol{\theta}_{q}^{\top} ,\boldsymbol{u}^{\top})^{\top}$ such that equation \eqref{eqn:matrixmod} becomes $\log(\boldsymbol{\mu}) = C_{\rho}\boldsymbol{\xi}$. Moreover, let  $\lambda_{spat} := \lambda_{q+1}$ and $\boldsymbol{\lambda} = (\lambda_1, \dots, \lambda_q, \lambda_{q+1})^{\top}$. Denote the precision of $\boldsymbol{\xi}$ by $Q_{\boldsymbol{\xi}}^{\boldsymbol{\lambda}} = \text{blkdiag}(V_{\beta}, \; \lambda_1 P, \; \dots, \;  \lambda_q P, 
 \; \lambda_{q+1} \Omega
 _{\rho})$, where $\text{blkdiag}(\cdot)$ is a block diagonal matrix. The full Bayesian model is given by:
 \begin{align*}
    & (y_i|\boldsymbol{\xi}) \sim \text{EDF}(\gamma_i, \phi), \quad i=1,\dots, n,\\
    & (\boldsymbol{\xi}|\boldsymbol{\lambda}, \rho)\sim \mathcal{N}_{dim(\boldsymbol{\xi})}(\boldsymbol{0},(Q_{\boldsymbol{\xi}}^{\boldsymbol{\lambda}})^{-1}),\\
    & (\lambda_j|\delta_j)\sim \mathcal{G}\left(\frac{\nu}{2},\frac{\nu \delta_j}{2}\right), \quad j=1,\dots, q + 1,\\
    & \delta_j \sim \mathcal{G}(a_\delta,b_\delta), \quad j=1,\dots, q + 1,\\
    & p(\rho) \propto \rho^{-1}, \\
    & p(\phi) \propto \phi^{-1},
\end{align*}

where  $\mathcal{G}(a,b)$ denotes a Gamma distribution with mean $a/b$ and variance $a/b^2$. This robust prior specification on the penalty parameters follows from \cite{jullion2007} where $a_\delta = b_\delta$ are chosen to be small enough, say $10^{-5}$, with fixed $\nu$  (e.g. $\nu=3$ in this paper). 

\subsection{Laplace approximation}

This section discusses the derivations for the conditional posterior distribution of $\boldsymbol{\xi}$ and approximate posterior distributions of the hyperparameters. The Laplace approximation is used to approximate the conditional posterior $p(\boldsymbol{\xi}|\boldsymbol{\lambda}, \rho, \phi,\mathcal{D})$ as a Gaussian distribution. This posterior approximation is particularly advantageous for its computational efficiency, significantly reducing computational time, as it eliminates the need for sampling compared to MCMC methods. In the case of a Gaussian response, the derived conditional posterior is exactly Gaussian, and the detailed derivations are provided in Appendix A. Denote the (Poisson or negative binomial) likelihood function by $\mathcal{L}(\boldsymbol{\xi}, \rho, \phi ;\mathcal{D})$ where $\mathcal{D}$ is the observed data. Using Bayes' rule, the conditional posterior of $\boldsymbol{\xi}$ can be written as  $ p(\boldsymbol{\xi}|\boldsymbol{\lambda}, \rho, \phi,\mathcal{D}) \propto \mathcal{L}(\boldsymbol{\xi}, \rho, \phi;\mathcal{D})p(\boldsymbol{\xi}|\boldsymbol{\lambda}, \rho)$. The gradient and Hessian of the log-conditional posterior, $\log p(\boldsymbol{\xi}|\boldsymbol{\lambda}, \rho, \phi,\mathcal{D})$, with respect to $\boldsymbol{\xi}$ are analytically derived and used in a Newton-Raphson algorithm to approximate the mode of the conditional posterior of $\boldsymbol{\xi}$. The availability of the derived analytic gradient and Hessian further enhances the computational speed. After convergence, the Laplace approximation of $p(\boldsymbol{\xi}|\boldsymbol{\lambda}, \rho, \phi,\mathcal{D})$ is a multivariate Gaussian density denoted by $\widetilde{p}_G(\boldsymbol{\xi}|\boldsymbol{\lambda},\rho, \phi, \mathcal{D})=\mathcal{N}_{\text{dim}(\boldsymbol{\xi})}(\widehat{\boldsymbol{\xi}}_{\boldsymbol{\lambda}},\widehat{\Sigma}_{\boldsymbol{\lambda}})$ where $\widehat{\boldsymbol{\xi}}_{\boldsymbol{\lambda}}$ is the mode and $\widehat{\Sigma}_{\boldsymbol{\lambda}}$ is the inverse of the negative Hessian matrix evaluated at the posterior mode.\\

Next, the (approximate) joint posterior distribution of the hyperparameters $\boldsymbol{\lambda}$, $\boldsymbol{\delta}$, $\rho$ and $\phi$ is derived. Let $\delta = (\delta_1, \delta_2, \dots, \delta_q)^{\top}$. Using Bayes' theorem, the joint marginal posterior of $\boldsymbol{\lambda}$, $\boldsymbol{\delta}$, $\rho$ and $\phi$ can be written as: 
$$p(\boldsymbol{\lambda}, \boldsymbol{\delta}, \rho, \phi|\mathcal{D}) \propto \frac{\mathcal{L}(\boldsymbol{\xi}, \rho, \phi;\mathcal{D})p(\boldsymbol{\xi}|\boldsymbol{\lambda})p(\boldsymbol{\lambda}|\boldsymbol{\delta})p(\boldsymbol{\delta})p(\rho)p(\phi)}{p(\boldsymbol{\xi}|\boldsymbol{\lambda}, \rho, \phi, \mathcal{D})}.$$
Following \cite{rue2009}, this joint posterior can be approximated by replacing the denominator $p(\boldsymbol{\xi}|\boldsymbol{\lambda}, \rho, \phi, \mathcal{D})$ with $\widetilde{p}_G(\boldsymbol{\xi}|\boldsymbol{\lambda},\rho, \phi, \mathcal{D})$ and by evaluating $\boldsymbol{\xi}$ at $\widehat{\boldsymbol{\xi}}_{\boldsymbol{\lambda}}$. Note that the determinant $|Q_{\boldsymbol{\xi}}^{\boldsymbol{\lambda}}|^{\frac{1}{2}}$ in $ p({\boldsymbol{\xi}}|\boldsymbol{\lambda})$ can be obtained as $|Q_{\boldsymbol{\xi}}^{\boldsymbol{\lambda}}|^{\frac{1}{2}} = \Big( |V_{\beta}| \times |\lambda_1 P| \times \cdots \times |\lambda_q P| \times |\lambda_{q+1} \Omega_{\rho}| \Big)^{\frac{1}{2}} \propto \Big( \lambda_1^{K} \times \cdots \times \lambda_q^{K} \times \lambda_{q+1}^{S}|\Omega_{\rho}| \Big) ^{\frac{1}{2}}$. The approximate joint posterior can then be written as:
\begin{align}
     \widetilde{p}(\boldsymbol{\lambda}, \boldsymbol{\delta}, \rho, \phi|\mathcal{D})  
     & \propto \mathcal{L}(\boldsymbol{\xi}, \rho, \phi;\mathcal{D}) \times \exp\left(- \frac{1}{2}\widehat{\boldsymbol{\xi}}_{\boldsymbol{\lambda}}^{\top} Q_{\boldsymbol{\xi}}^{\boldsymbol{\lambda}} \widehat{\boldsymbol{\xi}}_{\boldsymbol{\lambda}}\right) \times \prod_{j=1}^{q}(\lambda_j)^{\frac{K + \nu}{2} - 1} \times \lambda_{q+1}^{\frac{S + \nu}{2} - 1} \nonumber \\
     & \quad \times \prod_{j=1}^{q+1}(\delta_j)^{\frac{\nu}{2} + a_{\delta}-1} \exp\left(-\left(\frac{\nu \lambda_j}{2} + b_{\delta}\right)\delta_j\right)  \nonumber \\
     & \quad \times |\Omega_{\rho}|^{\frac{1}{2}} \rho^{-1} \phi^{-1}  |\widehat{\Sigma}_{\boldsymbol{\lambda}}|^{\frac{1}{2}}.
     \label{eqn:jointhypCount}
\end{align}
From \eqref{eqn:jointhypCount}, the joint posterior $\widetilde{p}(\boldsymbol{\lambda},\rho, \phi|\mathcal{D})$ can be analytically obtained by integrating out the hyperparameters $\delta_j$. To ensure numerical stability, the remaining hyperparameters are log-transformed such that $\boldsymbol{v}=(v_1, \dots, v_{q+1})^{\top}=(\log(\lambda_1),\dots,\log(\lambda_{q+1}))^{\top}$, $v_{\rho}=\log(\rho)$ and $v_{\phi}=\log(\phi)$. Note that the transformed posterior is multiplied by the Jacobian of the transformation given by $J = \prod_{j=1}^{q+1}\exp(v_j) \times \exp(v_\rho) \times \exp(v_\phi)$. The joint log-posterior of $\boldsymbol{v}$, $v_{\rho}$ and $v_{\phi}$  is then given by:
\begin{align*}
    \log \widetilde{p}(\boldsymbol{v}, v_{\phi} | D) & \dot{=} \log \mathcal{L}(\boldsymbol{\xi}, v_{\rho}, v_{\phi};\mathcal{D}) -0.5\widehat{\boldsymbol{\xi}}_{\boldsymbol{v}}^{\top} Q_{\boldsymbol{\xi}}^{\boldsymbol{v}} \widehat{\boldsymbol{\xi}}_{\boldsymbol{v}} + \sum_{j=1}^{q}\left(\frac{K + \nu}{2} v_j\right) + \left(\frac{S + \nu}{2} v_{q+1}\right) \\ 
    & \quad - \sum_{j=1}^{q+1} \left(\frac{\nu}{2} + a_{\delta}\right) \log \left( \frac{\nu \exp(v_j)}{2} + b_{\delta}\right) + \frac{1}{2} \log|\Omega_{v_{\rho}}| + \frac{1}{2} \log |\widehat{\Sigma}_{\boldsymbol{v}}|,
\end{align*}
where $\dot{=}$ denotes equality up to an additive constant. The above log-posterior is then optimized to obtain the maximum a posteriori estimate for  $\boldsymbol{v}$, $v_{\rho}$ and $v_{\phi}$.

\subsection{Prediction and prediction interval}
The main goal of kriging is to predict the value of the response variable $y_0$ at an arbitrary location $\boldsymbol{w}_0 = (w_{1_{0}}, w_{2_{0}})^{\top}$. Suppose the linear and smooth covariates are available denoted by $(x_{1_{0}}, x_{2_{0}}, \dots, x_{p_{0}})$ and $(s_{1_{0}}, s_{2_{0}}, \dots, s_{q_{0}})$, respectively. Define $\boldsymbol{x}_{0} = (1, x_{1_{0}}, x_{2_{0}}, \dots, x_{p_{0}}, w_{1_{0}}, w_{2_{0}})^{\top}$ and denote the B-spline basis for the smooth covariates by $\boldsymbol{b}_{j_{0}} = (b_{j1}(s_{j_{0}}), b_{j2}(s_{j_{0}}), \dots, b_{jK}(s_{j_{0}}))^{\top}$ for $j = 1, \dots, q$. The spline basis for the coordinates is given by $\boldsymbol{z}_{0}(\widehat{\rho}) = (z_{1_{0}}(\widehat{\rho}), z_{2_{0}}(\widehat{\rho}), \dots, z_{S_{0}}(\widehat{\rho}))^{\top}$, where $z_{s_{0}}(\widehat{\rho}) = R_{\widehat{\rho}}(\boldsymbol{w}_0 - \boldsymbol{\kappa}_s)$ for $s = 1, \dots, S$ and $\widehat{\rho}$ is the maximum a posteriori estimate of $\rho$. Furthermore, define $\boldsymbol{c}_{\widehat{\rho}} = (\boldsymbol{x}_0^{\top}, \boldsymbol{b}_{1_{0}}^{\top}, \boldsymbol{b}_{2_{0}}^{\top}, \dots, \boldsymbol{b}_{q_{0}}^{\top}, \boldsymbol{z}_{0}(\widehat{\rho})^{\top})^{\top}$. From model \eqref{eqn:matrixmod}, the estimated mean response can be obtained as $\widehat{\mathbb{E}(y_0)} = \exp(\boldsymbol{c}_{\widehat{\rho}}^{\top} \widehat{\boldsymbol{\xi}}_{\boldsymbol{\lambda}})$.\\

To obtain predictions, note that $\log(\mathbb{E}(y_0)) = \boldsymbol{c}_{\rho}^{\top} \boldsymbol{\xi}$, where $\widetilde{p}_G(\boldsymbol{\xi} | \boldsymbol{v}, v_\rho, v_\phi, \mathcal{D}) = \mathcal{N}_{\text{dim}(\boldsymbol{\xi})}(\widehat{\boldsymbol{\xi}}_{\boldsymbol{v}}, \widehat{\Sigma}_{\boldsymbol{v}})$. The approximate posterior distribution for the log mean number of cases is thus $\widetilde{p}(\log(\mathbb{E}(y_0)) | \boldsymbol{v}, v_\rho, v_\phi, \mathcal{D}) = \mathcal{N}_1(\boldsymbol{c}_{\widehat{\rho}}^{\top} \widehat{\boldsymbol{\xi}}_{\boldsymbol{v}}, \boldsymbol{c}_{\widehat{\rho}}^{\top} \widehat{\boldsymbol{\Sigma}}_{\boldsymbol{v}} \boldsymbol{c}_{\widehat{\rho}})$. From this Gaussian distribution, 1000 samples are drawn with mean $\boldsymbol{c}_{\widehat{\rho}}^{\top} \widehat{\boldsymbol{\xi}}_{\boldsymbol{v}}$ and variance $\boldsymbol{c}_{\widehat{\rho}}^{\top} \widehat{\boldsymbol{\Sigma}}_{\boldsymbol{v}} \boldsymbol{c}_{\widehat{\rho}}$. These samples are exponentiated to obtain an estimated mean vector, denoted by $\boldsymbol{\mu}_0$. Subsequently, 1000 samples are generated from a Poisson or negative binomial distribution with mean vector $\boldsymbol{\mu}_0$ to obtain predictive samples for $y_0$. The corresponding quantiles of these samples are then computed to determine the desired prediction interval. Note that in the presence of an offset term $N_i$, both $\widehat{y}_0$ and $\boldsymbol{\mu}_0$ are multiplied by $N_i$.

\subsection{Bayesian information 
criteria and test for smooth effects}
The effective degrees of freedom (ED) serve as a measure of model complexity and quantify the amount of smoothing in a given model. A covariate with an associated ED value close to 1 indicates a linear effect, while values greater than 1 indicate nonlinearity. Let $\mathcal{I} = -\nabla^2_{\boldsymbol{\xi}}\log \mathcal{L}(\boldsymbol{\xi}, \rho, \phi;\mathcal{D})\bigr|_{\bm{\xi}=\widehat{\boldsymbol{\xi}}_{\boldsymbol{\lambda}}}$ denote the negative Hessian of the log-likelihood evaluated at the posterior mode estimate $\widehat{\boldsymbol{\xi}}_{\boldsymbol{\lambda}}$. The total ED is computed by summing the main diagonal of the matrix $\mathcal{H} =\widehat{\Sigma}_{\boldsymbol{\lambda}}\mathcal{I}$, where $\widehat{\Sigma}_{\boldsymbol{\lambda}}$ is the estimated covariance matrix of $\boldsymbol{\xi}$. The ED of a specific smooth term is obtained by summing the diagonal elements of $\mathcal{H}$ that correspond to the B-spline coefficients associated with the smooth term. The Bayesian information criteria (BIC) \citep{schwarz1978} is useful for model selection and is computed using the formula $BIC = -2\log \mathcal{L}(\boldsymbol{\xi}, \rho, \phi;\mathcal{D}) + \text{ED}\times \log(n)$. A lower BIC indicates a better fit of the model.\\

\cite{wood2013} proposed a test statistic, $T_r$, to test the effect of a smooth covariate. $T_r$ is a Wald-type statistic used to test the null hypothesis $H_0: f_j(x) = 0$ versus the alternative hypothesis  $H_a: f_j(x) \neq  0$ for a smooth covariate $x$. Let $\widehat{f}_j = B_j \widehat{\boldsymbol{\theta}}_j $ denote the estimated \textit{j}th smooth function, where $B_j$ and $\widehat{\boldsymbol{\theta}}_j$ are the associated B-spline matrix and estimated coefficients, respectively. The covariance matrix of $\widehat{f}_j$ is given by $V_{\widehat{f}_j} = B_j\widehat{\Sigma}_{\boldsymbol{\theta}_j}B^{\top}_j$, where $\widehat{\Sigma}_{\boldsymbol{\theta}_j}$ is the estimated covariance matrix of the B-spline coefficients $\boldsymbol{\theta}_j$. The test statistic is computed as $T_r = \widehat{f}^{\top}_jV^{r-}_{\widehat{f}_j}\widehat{f}_j$, where $V^{r-}_{\widehat{f}_j}$ is the
rank-\textit{r} Moore-Penrose inverse of $V_{\widehat{f}_j}$ and \textit{r} is the estimated
effective degrees of freedom for the \textit{j}th smooth. Under the null hypothesis, $T_r$ follows a Gamma distribution, that is $T_r \sim \mathcal{G}(r/2, 1/2)$, with mean $\mathbb{E}(T_r) = r$ and variance $\mathbb{V}(T_r) = 2r$.

\section{Simulation study}
\begin{figure}[H]
\centering
\subfloat[Function $s_1$]{\includegraphics[width=.33\linewidth]{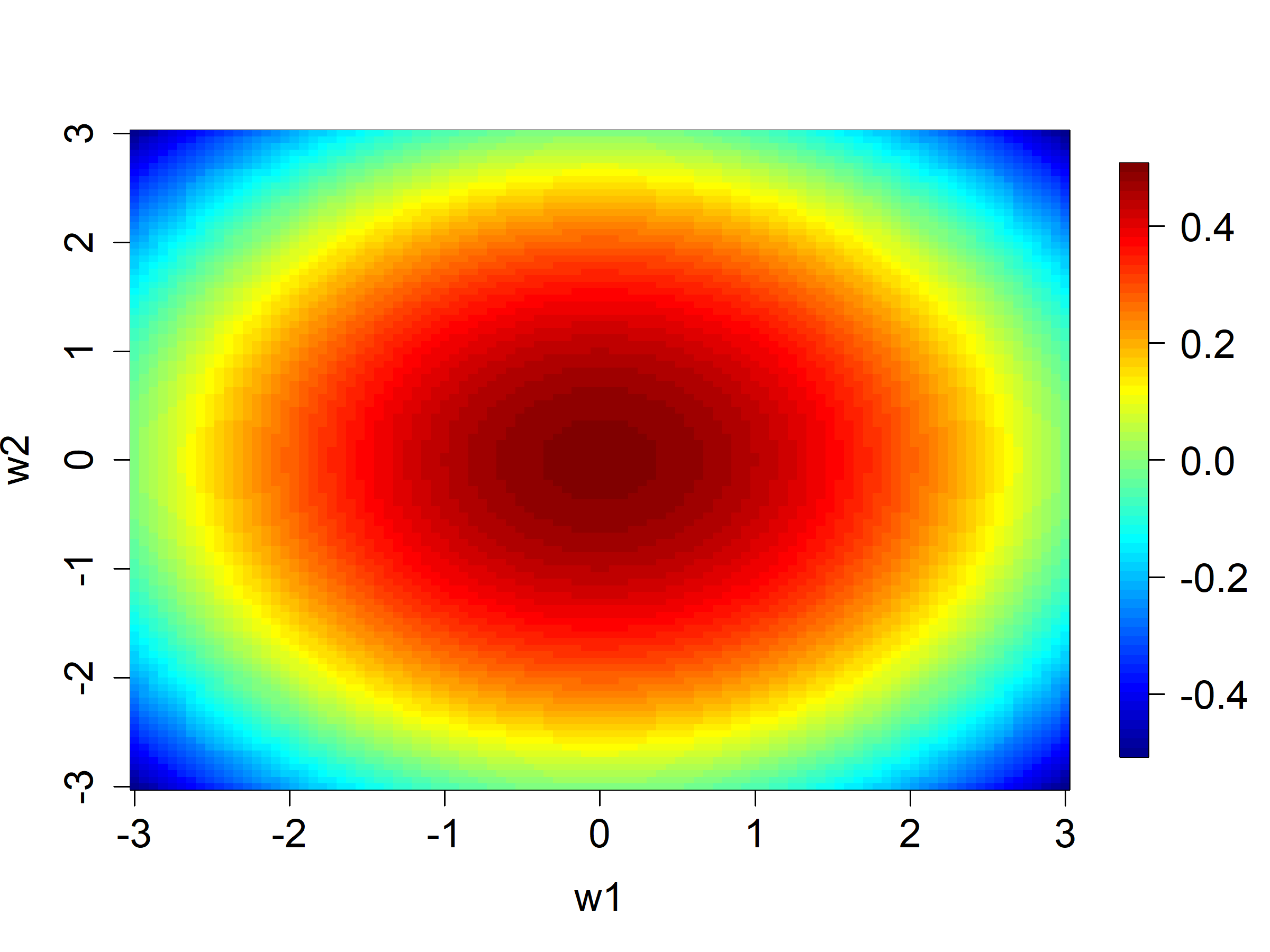}}
\subfloat[Function $s_2$]{\includegraphics[width=.33\linewidth]{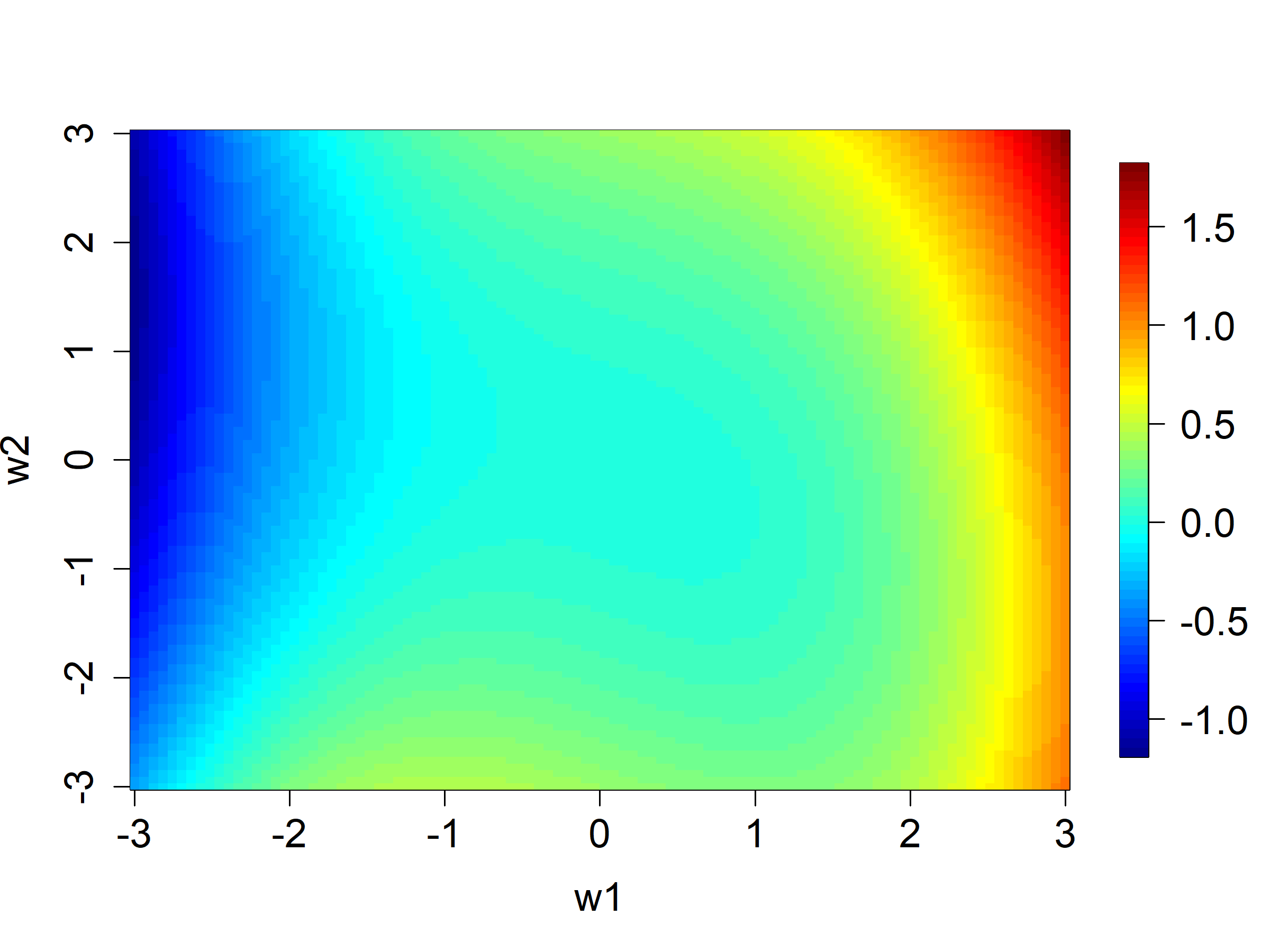}}\
\subfloat[Function $s_3$]{\includegraphics[width=.33\linewidth]{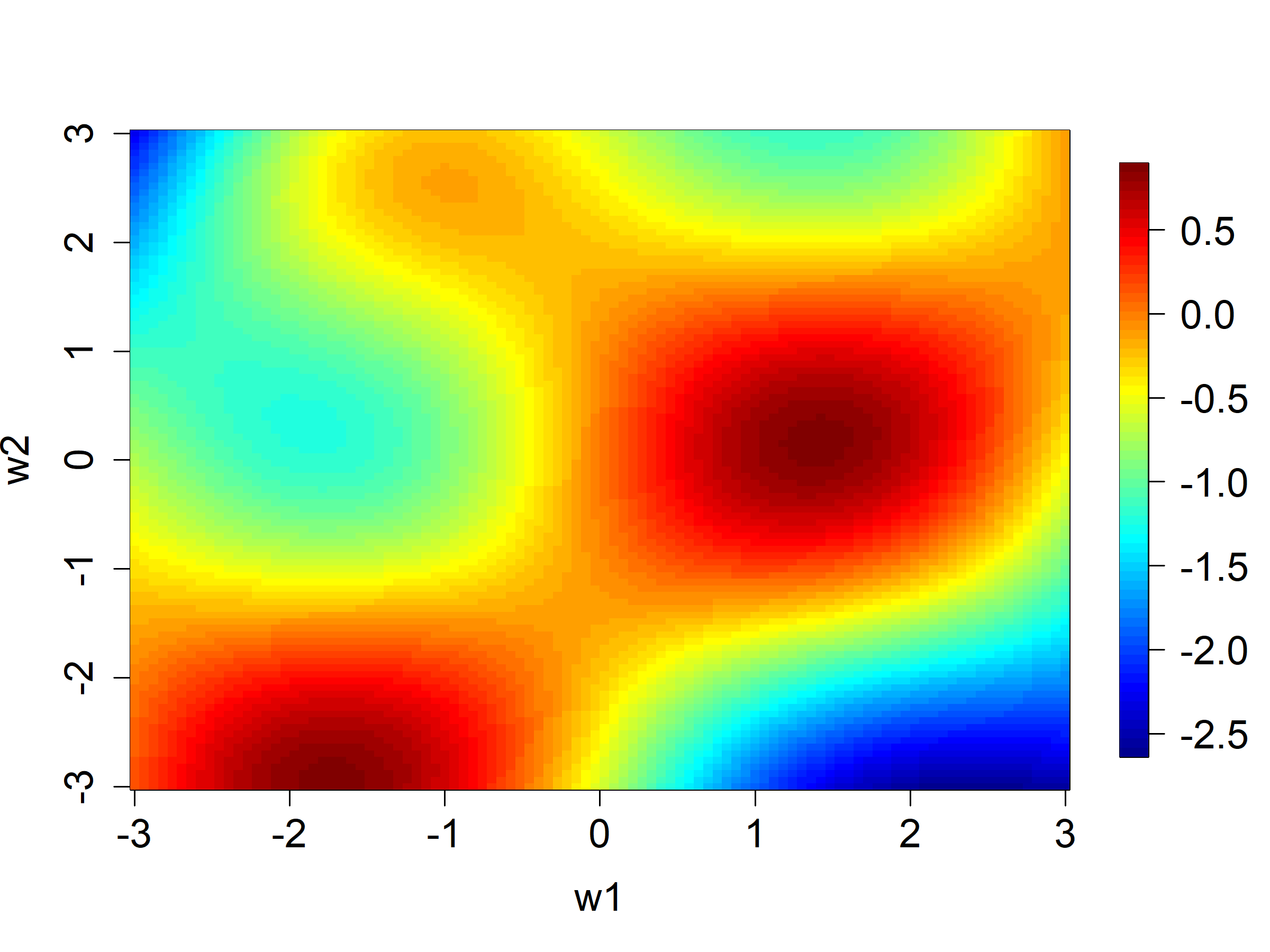}}
\caption{Two-dimensional smooth functions.}
\label{fig:2dfunc}
\end{figure}

A simulation study is conducted to evaluate the performance of our proposed methodology. For the count data, the samples $y$ are generated from a Poisson distribution with rate parameter $\mu \cdot \exp(\varepsilon)$. The inclusion of $\exp(\varepsilon)$ preserves the stochasticity of the true spatial process in agreement with equation \eqref{eqn:lowrankspat}. Thus, this is equivalent to simulating a spatial component from a two-dimensional smooth function plus an error term $(s(w_1, w_2) + \varepsilon)$. For the Gaussian data, $\varepsilon$ represents the error term of the Gaussian distribution, such that observations $y$ have mean $\mu$ and variance $\mathbb{V}(\varepsilon) = \sigma^2$. In the simulation, the following mean structure is assumed:
\[
\mu = \beta_0 + \beta_1 x_1 + f(x_2) + s(w_1, w_2),
\]
where $\beta_0 = 3$, $\beta_1 = -0.5$, and $f(x_2) = \cos(2 \pi x_2)$. The covariates $x_1$ and $x_2$ are randomly simulated from a uniform distribution over the unit interval, and $\varepsilon$ is drawn from a zero mean Gaussian distribution with standard deviation $\sigma = 0.25$ for the count data and $\sigma = \sqrt{0.10}$ for the Gaussian data. The spatial component $s(w_1, w_2)$ is a two-dimensional smooth function, for which the following three different forms are considered:

\[
\begin{aligned}
& s_1(w_1, w_2) = 0.5 - \frac{w_1^2 + w_2^2}{18}, \\
& s_2(w_1, w_2) = \frac{w_1^3 + w_1 w_2 + w_2^2}{25}, \\
& s_3(w_1, w_2) = -\frac{(w_1 - w_2)^2}{15} + \sin(w_1)\cos(w_2).
\end{aligned}
\]
Here, the spatial coordinates $w_1$ and $w_2$ are simulated from a uniform distribution on the interval $(-3, 3)$. The plot for the two-dimensional smooth functions considered in the simulation is shown in Figure \ref{fig:2dfunc}.

The performance measures used in our study are bias, relative bias (\%Bias), and credible interval (CI) coverage. These measures are computed for $\mu$, the smooth term $f(x_2)$, and the spatial term $s(w_1, w_2)$. Additionally, the prediction interval (PI) coverage for the samples $y$ are calculated. For smooth and spatial terms, a grid of values for $x_2$ and a two-dimensional grid for $(w_1, w_2)$ are created. Using these grids, $f(x_2)$ and $s(w_1, w_2)$ are computed, and the corresponding relative bias and PI coverage are obtained. To avoid identifiability issues, column mean centering is imposed for the basis functions associated with the smooth and spatial components and the performance measures are computed on the mean centered $f(x_2)$ and $s(w_1, w_2)$. Due to this centering, the bias is only computed for $\mu$. Suppose that the quantity of interest is denoted by $\omega$ with the corresponding estimate $\widehat{\omega}$. The bias and relative bias are computed as follows:

\[
\text{Bias} = \frac{1}{B \times M} \sum_{b=1}^{B} \sum_{m=1}^{M} (\omega_{bm} - \widehat{\omega}_{bm}),
\]

\[
\% \text{Bias} = \frac{1}{B \times M} \sum_{b=1}^{B} \sum_{m=1}^{M} \left| \frac{\omega_{bm} - \widehat{\omega}_{bm}}{\omega_{bm}} \right| \times 100\%,
\]

where $M$ is the length of the grid created for the covariates, and $B$ is the number of simulations or generated realizations. Moreover, CI or PI coverage is determined by calculating the percentage of $\omega$ values that fall within the interval.\\

Tables \ref{tab:results_s1}-\ref{tab:results_s3} present the results for different functions ($s_1$, $s_2$ and $s_3$) across various covariance structures (circular, exponential, Mat\'ern, and spherical) and different model distributions (negative binomial, Poisson, and Gaussian). The percentage bias for the smooth term is around 4\% to 7\% for all scenarios (Tables 1-3), demonstrating low bias and suggesting robust performance in estimating smooth effects using our proposed method. For all functions ($s_1, s_2 \; \text{and} \; s_3$), the negative binomial model has coverage rates of the smooth term that are generally closer to the 95\% nominal level for the smooth term as compared to the Poisson model which shows undercoverage, with rates ranging from 73\% to 80\%. The Gaussian data, on the other hand, also demonstrates good coverage, ranging from 91\% to 94\%, which is slightly lower than the nominal level. Note that the relative bias for the Gaussian data is very low, around 4\% to 5\%.\\

Regarding the spatial term, the Gaussian model generally exhibits the lowest relative bias, followed by the negative binomial model, while the Poisson model has the highest bias. Notably, all covariance functions show similar relative biases across each model for the spatial term, except for the Mat\'ern covariance function. For the Poisson model, the Mat\'ern covariance consistently shows lower bias. In the negative binomial and Gaussian models, the Mat\'ern covariance has a relatively low bias for function $s_1$ but a higher bias for functions $s_2$ and $s_3$. The coverage for the spatial term is highest for the negative binomial model, ranging from 97\% to 99\%, except for the Mat\'ern covariance, which has lower coverage for functions $s_2$ and $s_3$, with rates of 83\% and 69\%, respectively. This pattern is similar to the Gaussian data, with coverage mostly close to the 95\% nominal level, although there is undercoverage for the Mat\'ern covariance for functions $s_2$ and $s_3$. This behavior for the spatial term suggests that using our proposed model, the Mat\'ern covariance might be more suitable for symmetric functions ($s_1$) but not for non-symmetric functions in modeling the spatial component. On the other hand, the other covariance functions are more robust regardless of the shape of the underlying spatial structure. Finally, the Poisson model shows undercoverage for all functions and covariance structures.\\

For the mean response, biases are generally close to zero for all scenarios, with the percentage bias being lowest for the Gaussian data, ranging between 1\% to 3\%, except for the Mat\'ern covariance with function $s_3$, where the relative bias is around 6\%. The relative bias for $\mu$ is mostly lower for the negative binomial model compared to the Poisson model, but their values are very similar, ranging around 4\% to 9\%, except for function $s_3$ with the Mat\'ern covariance in the negative binomial model, which has a relative bias of 13\%. Finally, for the response $y$, the Gaussian model has prediction interval coverage very close to the nominal level. For count data, the Poisson model shows undercoverage, while the negative binomial model has high prediction interval coverage. Overall, the negative binomial and Gaussian models demonstrate robust performance, characterized by low bias and high coverage rates across all quantities of interest. In contrast, although the Poisson model exhibits low bias, it suffers from lower coverage rates due to the presence of an additional error term, treated as an overdispersion parameter, that is unaccounted for within the Poisson model.\\

Additional simulation results without covariates are presented in Appendix B, comparing the proposed model with the classical kriging approach for the Gaussian data using the \texttt{geoR} package, yielding comparable results, although the low-rank approach has a slightly larger percentage bias in terms of the mean. Moreover, the computation time is compared using the \texttt{microbenchmark()} function from the \texttt{microbenchmark} package in R with 10 functions evaluated for 1000 observations. The analysis is implemented on a device with an Intel(R) Core(TM) i5-1135G7, CPU running at a base frequency of 2.40GHz, and having 4 cores with 16GB of RAM.
The average real elapsed time for the proposed Bayesian approach is around 1 second, while the classical kriging approach takes around 110 seconds on average  (see Table B.2 in the Appendix). This highlights the computational benefit of our methodology.

\begin{table}[H]
\centering
\caption{Simulation results for function $s_1$. Model - distributional assumption for the response.  Covariance - covariance functions used in modeling the spatial component. CP(\%) - indicates 95\% interval coverage probability.}
\label{tab:results_s1}
\resizebox{\textwidth}{!}{
\begin{tabular}{|l|l|cc|cc|cc|c|}
\hline
Model & Covariance & \multicolumn{2}{c|}{Smooth term $f(x_2)$}        & \multicolumn{2}{c|}{Spatial term $s(w_1,w_2)$}       & \multicolumn{2}{c|}{$\mu$}             & $y$     \\ \hline
\multicolumn{1}{|c|}{}          & \multicolumn{1}{c|}{} & \multicolumn{1}{c|}{Bias(\%)} & CP(\%) & \multicolumn{1}{c|}{Bias(\%)} & CP(\%) & \multicolumn{1}{c|}{Bias}  & Bias (\%) & CP(\%) \\ \hline
\multirow{4}{*}{\shortstack{Negative \\ binomial}}             & Circular              & \multicolumn{1}{c|}{5.49}     & 94.08   & \multicolumn{1}{c|}{23.43}    & 99.43   & \multicolumn{1}{c|}{-0.96} & 5.42      & 98.11   \\ \cline{2-9} 
                                & Exponential           & \multicolumn{1}{c|}{5.41}     & 93.94   & \multicolumn{1}{c|}{22.95}    & 98.99   & \multicolumn{1}{c|}{-1.09} & 5.39      & 98.16   \\ \cline{2-9} 
                                & Mat\'ern                & \multicolumn{1}{c|}{5.40}     & 93.66   & \multicolumn{1}{c|}{17.77}    & 98.36   & \multicolumn{1}{c|}{-0.86} & 4.71      & 98.10   \\ \cline{2-9} 
                                & Spherical             & \multicolumn{1}{c|}{5.32}     & 93.62   & \multicolumn{1}{c|}{23.24}    & 99.25   & \multicolumn{1}{c|}{-1.00} & 5.38      & 98.10   \\ \hline
                                &                       & \multicolumn{1}{c|}{}         &         & \multicolumn{1}{c|}{}         &         & \multicolumn{1}{c|}{}      &           &         \\ \hline
\multirow{4}{*}{Poisson}        & Circular              & \multicolumn{1}{c|}{5.92}     & 76.84   & \multicolumn{1}{c|}{39.35}    & 80.73   & \multicolumn{1}{c|}{-0.91} & 7.57      & 83.00   \\ \cline{2-9} 
                                & Exponential           & \multicolumn{1}{c|}{5.80}     & 76.39   & \multicolumn{1}{c|}{38.20}    & 80.83   & \multicolumn{1}{c|}{-0.99} & 7.36      & 82.74   \\ \cline{2-9} 
                                & Mat\'ern                & \multicolumn{1}{c|}{5.52}     & 77.37   & \multicolumn{1}{c|}{19.20}    & 77.93   & \multicolumn{1}{c|}{-0.95} & 5.19      & 82.75   \\ \cline{2-9} 
                                & Spherical             & \multicolumn{1}{c|}{5.98}     & 76.46   & \multicolumn{1}{c|}{39.99}    & 80.63   & \multicolumn{1}{c|}{-0.93} & 7.65      & 83.33   \\ \hline
                                &                       & \multicolumn{1}{c|}{}         &         & \multicolumn{1}{c|}{}         &         & \multicolumn{1}{c|}{}      &           &         \\ \hline
\multirow{4}{*}{Gaussian}       & Circular              & \multicolumn{1}{c|}{4.88}     & 92.34   & \multicolumn{1}{c|}{19.87}    & 97.77   & \multicolumn{1}{c|}{-0.02} & 1.49      & 94.42   \\ \cline{2-9} 
                                & Exponential           & \multicolumn{1}{c|}{4.92}     & 92.42   & \multicolumn{1}{c|}{21.01}    & 96.73   & \multicolumn{1}{c|}{-0.02} & 1.55      & 94.38   \\ \cline{2-9} 
                                & Mat\'ern                & \multicolumn{1}{c|}{4.81}     & 91.74   & \multicolumn{1}{c|}{16.21}    & 95.04   & \multicolumn{1}{c|}{-0.01} & 1.26      & 94.77   \\ \cline{2-9} 
                                & Spherical             & \multicolumn{1}{c|}{4.88}     & 92.26   & \multicolumn{1}{c|}{20.06}    & 97.49   & \multicolumn{1}{c|}{-0.02} & 1.49      & 94.46   \\ \hline
\end{tabular}}
\end{table}

\begin{table}[H]
\centering
\caption{Simulation results for function $s_2$. Model - distributional assumption for the response.  Covariance - covariance functions used in modeling the spatial component. CP(\%) - indicates 95\% interval coverage probability.}
\label{tab:results_s2}
\resizebox{\textwidth}{!}{
\begin{tabular}{|l|l|cc|cc|cc|c|}
\hline
       Model         &    Covariance         & \multicolumn{2}{c|}{Smooth term  $f(x_2)$}        & \multicolumn{2}{c|}{Spatial term  $s(w_1,w_2)$}       & \multicolumn{2}{c|}{$\mu$}             & $y$     \\ \hline
\multirow{5}{*}{\shortstack{Negative \\ binomial}}       &             & \multicolumn{1}{c|}{Bias(\%)} & CP(\%) & \multicolumn{1}{c|}{Bias(\%)} & CP(\%) & \multicolumn{1}{c|}{Bias}  & Bias (\%) & CP(\%) \\ \cline{2-9} 
                          & Circular    & \multicolumn{1}{c|}{5.53}     & 93.78   & \multicolumn{1}{c|}{18.51}    & 97.43   & \multicolumn{1}{c|}{0.37}  & 6.49      & 97.90   \\ \cline{2-9} 
                          & Exponential & \multicolumn{1}{c|}{5.62}     & 92.53   & \multicolumn{1}{c|}{18.44}    & 97.36   & \multicolumn{1}{c|}{0.51}  & 6.62      & 97.88   \\ \cline{2-9} 
                          & Mat\'ern      & \multicolumn{1}{c|}{5.43}     & 94.12   & \multicolumn{1}{c|}{20.60}    & 83.10   & \multicolumn{1}{c|}{0.38}  & 7.06      & 98.02   \\ \cline{2-9} 
                          & Spherical   & \multicolumn{1}{c|}{5.47}     & 94.11   & \multicolumn{1}{c|}{18.90}    & 97.97   & \multicolumn{1}{c|}{0.20}  & 6.52      & 97.90   \\ \hline
                          &             & \multicolumn{1}{c|}{}         &         & \multicolumn{1}{c|}{}         &         & \multicolumn{1}{c|}{}      &           &         \\ \hline
\multirow{4}{*}{Poisson}  & Circular    & \multicolumn{1}{c|}{5.82}     & 76.46   & \multicolumn{1}{c|}{27.88}    & 80.23   & \multicolumn{1}{c|}{-0.41} & 7.89      & 82.09   \\ \cline{2-9} 
                          & Exponential & \multicolumn{1}{c|}{5.89}     & 76.12   & \multicolumn{1}{c|}{27.01}    & 80.38   & \multicolumn{1}{c|}{-0.18} & 7.68      & 81.72   \\ \cline{2-9} 
                          & Mat\'ern      & \multicolumn{1}{c|}{5.99}     & 73.84   & \multicolumn{1}{c|}{15.94}    & 75.44   & \multicolumn{1}{c|}{-0.48} & 5.99      & 81.31   \\ \cline{2-9} 
                          & Spherical   & \multicolumn{1}{c|}{6.35}     & 73.98   & \multicolumn{1}{c|}{28.58}    & 79.93   & \multicolumn{1}{c|}{-0.40} & 8.01      & 81.92   \\ \hline
                          &             & \multicolumn{1}{c|}{}         &         & \multicolumn{1}{c|}{}         &         & \multicolumn{1}{c|}{}      &           &         \\ \hline
\multirow{4}{*}{Gaussian} & Circular    & \multicolumn{1}{c|}{4.89}     & 92.48   & \multicolumn{1}{c|}{16.78}    & 94.76   & \multicolumn{1}{c|}{0.00}  & 1.94      & 94.37   \\ \cline{2-9} 
                          & Exponential & \multicolumn{1}{c|}{4.99}     & 92.84   & \multicolumn{1}{c|}{16.87}    & 93.59   & \multicolumn{1}{c|}{0.00}  & 1.98      & 94.28   \\ \cline{2-9} 
                          & Mat\'ern      & \multicolumn{1}{c|}{4.84}     & 92.61   & \multicolumn{1}{c|}{26.34}    & 52.35   & \multicolumn{1}{c|}{0.00}  & 2.59      & 94.66   \\ \cline{2-9} 
                          & Spherical   & \multicolumn{1}{c|}{4.92}     & 92.60   & \multicolumn{1}{c|}{16.69}    & 94.41   & \multicolumn{1}{c|}{0.00}  & 1.93      & 94.38   \\ \hline
\end{tabular}}
\end{table}

\begin{table}[H]
\centering
\caption{Simulation results for function $s_3$.  Model - distributional assumption for the response.  Covariance - covariance functions used in modeling the spatial component. CP(\%) - indicates 95\% interval coverage probability.}
\label{tab:results_s3}
\resizebox{\textwidth}{!}{
\begin{tabular}{|l|l|cc|cc|cc|c|}
\hline
Model & Covariance & \multicolumn{2}{c|}{Smooth term  $f(x_2)$}        & \multicolumn{2}{c|}{Spatial term  $s(w_1,w_2)$}       & \multicolumn{2}{c|}{$\mu$}             & $y$     \\ \hline
\multicolumn{1}{|c|}{}          & \multicolumn{1}{c|}{} & \multicolumn{1}{c|}{Bias(\%)} & CP(\%) & \multicolumn{1}{c|}{Bias(\%)} & CP(\%) & \multicolumn{1}{c|}{Bias}  & Bias (\%) & CP(\%) \\ \hline
\multirow{4}{*}{\shortstack{Negative \\ binomial}}             & Circular              & \multicolumn{1}{c|}{6.06}     & 93.95   & \multicolumn{1}{c|}{13.44}    & 97.16   & \multicolumn{1}{c|}{-0.34} & 8.05      & 98.15   \\ \cline{2-9} 
                                & Exponential           & \multicolumn{1}{c|}{5.83}     & 94.62   & \multicolumn{1}{c|}{13.60}    & 97.17   & \multicolumn{1}{c|}{-0.33} & 8.03      & 98.25   \\ \cline{2-9} 
                                & Mat\'ern                & \multicolumn{1}{c|}{7.31}     & 91.99   & \multicolumn{1}{c|}{19.86}    & 69.01   & \multicolumn{1}{c|}{-0.20} & 13.09     & 99.15   \\ \cline{2-9} 
                                & Spherical             & \multicolumn{1}{c|}{5.95}     & 94.93   & \multicolumn{1}{c|}{13.28}    & 97.60   & \multicolumn{1}{c|}{-0.34} & 7.97      & 98.26   \\ \hline
                                &                       & \multicolumn{1}{c|}{}         &         & \multicolumn{1}{c|}{}         &         & \multicolumn{1}{c|}{}      &           &         \\ \hline
\multirow{4}{*}{Poisson}        & Circular              & \multicolumn{1}{c|}{6.51}     & 79.43   & \multicolumn{1}{c|}{15.82}    & 83.74   & \multicolumn{1}{c|}{-0.53} & 8.92      & 87.19   \\ \cline{2-9} 
                                & Exponential           & \multicolumn{1}{c|}{6.46}     & 79.82   & \multicolumn{1}{c|}{15.84}    & 82.86   & \multicolumn{1}{c|}{-0.51} & 9.03      & 87.29   \\ \cline{2-9} 
                                & Mat\'ern                & \multicolumn{1}{c|}{6.22}     & 78.44   & \multicolumn{1}{c|}{12.75}    & 70.28   & \multicolumn{1}{c|}{-0.64} & 8.26      & 86.65   \\ \cline{2-9} 
                                & Spherical             & \multicolumn{1}{c|}{6.28}     & 79.42   & \multicolumn{1}{c|}{15.85}    & 83.77   & \multicolumn{1}{c|}{-0.45} & 8.98      & 87.36   \\ \hline
                                &                       & \multicolumn{1}{c|}{}         &         & \multicolumn{1}{c|}{}         &         & \multicolumn{1}{c|}{}      &           &         \\ \hline
\multirow{4}{*}{Gaussian}       & Circular              & \multicolumn{1}{c|}{5.01}     & 92.66   & \multicolumn{1}{c|}{11.10}    & 95.28   & \multicolumn{1}{c|}{-0.01} & 2.85      & 94.70   \\ \cline{2-9} 
                                & Exponential           & \multicolumn{1}{c|}{4.97}     & 92.41   & \multicolumn{1}{c|}{11.24}    & 94.76   & \multicolumn{1}{c|}{-0.02} & 2.87      & 94.64   \\ \cline{2-9} 
                                & Mat\'ern                & \multicolumn{1}{c|}{5.32}     & 94.14   & \multicolumn{1}{c|}{22.74}    & 44.49   & \multicolumn{1}{c|}{-0.02} & 5.91      & 96.31   \\ \cline{2-9} 
                                & Spherical             & \multicolumn{1}{c|}{5.00}     & 93.25   & \multicolumn{1}{c|}{11.23}    & 95.17   & \multicolumn{1}{c|}{-0.02} & 2.87      & 94.55   \\ \hline
\end{tabular}}
\end{table}

\section{Data application}

In this section, the proposed methodology is applied to the analysis of two datasets: (1) Meuse river data using the Gaussian model and (2) vulnerability to coronavirus disease 2019 (COVID-19) in Belgium using the negative binomial model.

\subsection{Meuse river data}

The Meuse dataset contains measurements of heavy metal concentrations in topsoil collected from the flood plain of the Meuse River near the village of Stein, Netherlands. It also includes the geographic coordinates of each sampling location and is commonly used to demonstrate kriging and other geostatistical techniques. This analysis focuses on zinc concentrations in the topsoil, using two covariates: (1) distance to the Meuse river (dist), and (2) relative elevation above the local riverbed (elev). The dataset is available in the R package \texttt{sp}.

\begin{figure}[H]
\centering
\subfloat[]{\label{fig:obsMeusea}\includegraphics[width=.33\linewidth]{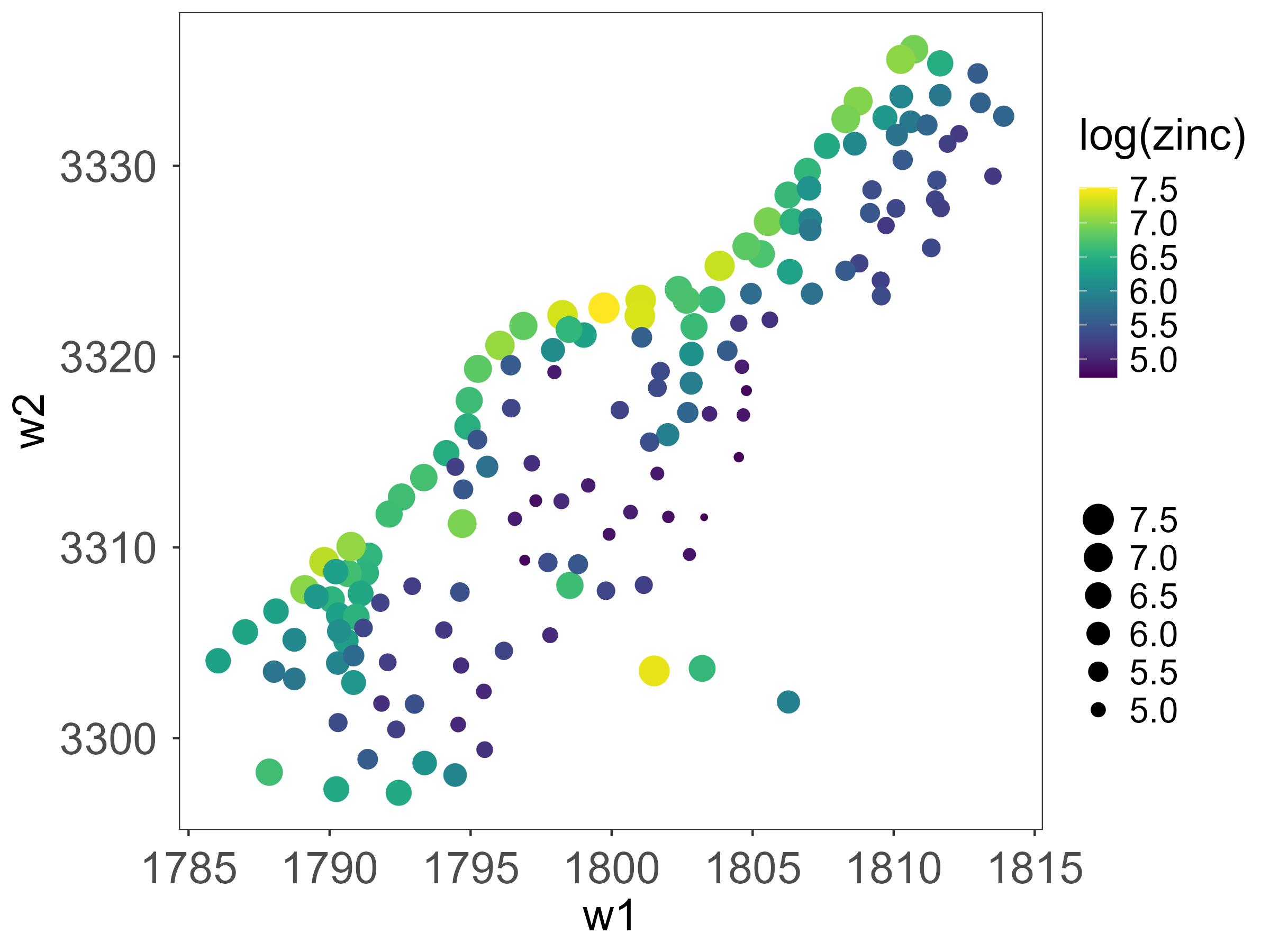}}
\subfloat[]{\label{fig:obsMeuseb}\includegraphics[width=.33\linewidth]{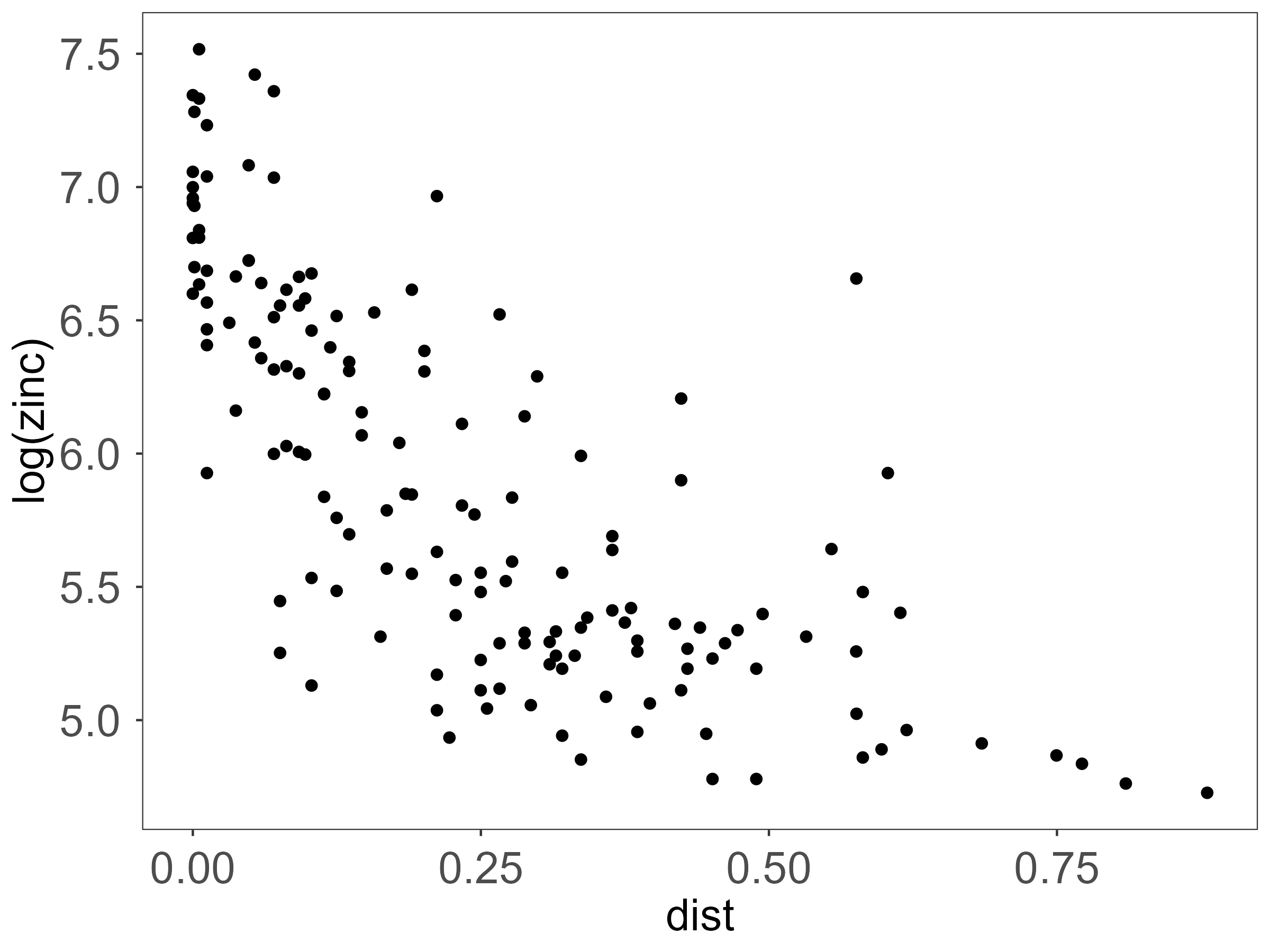}}\
\subfloat[]{\label{fig:obsMeusec}\includegraphics[width=.33\linewidth]{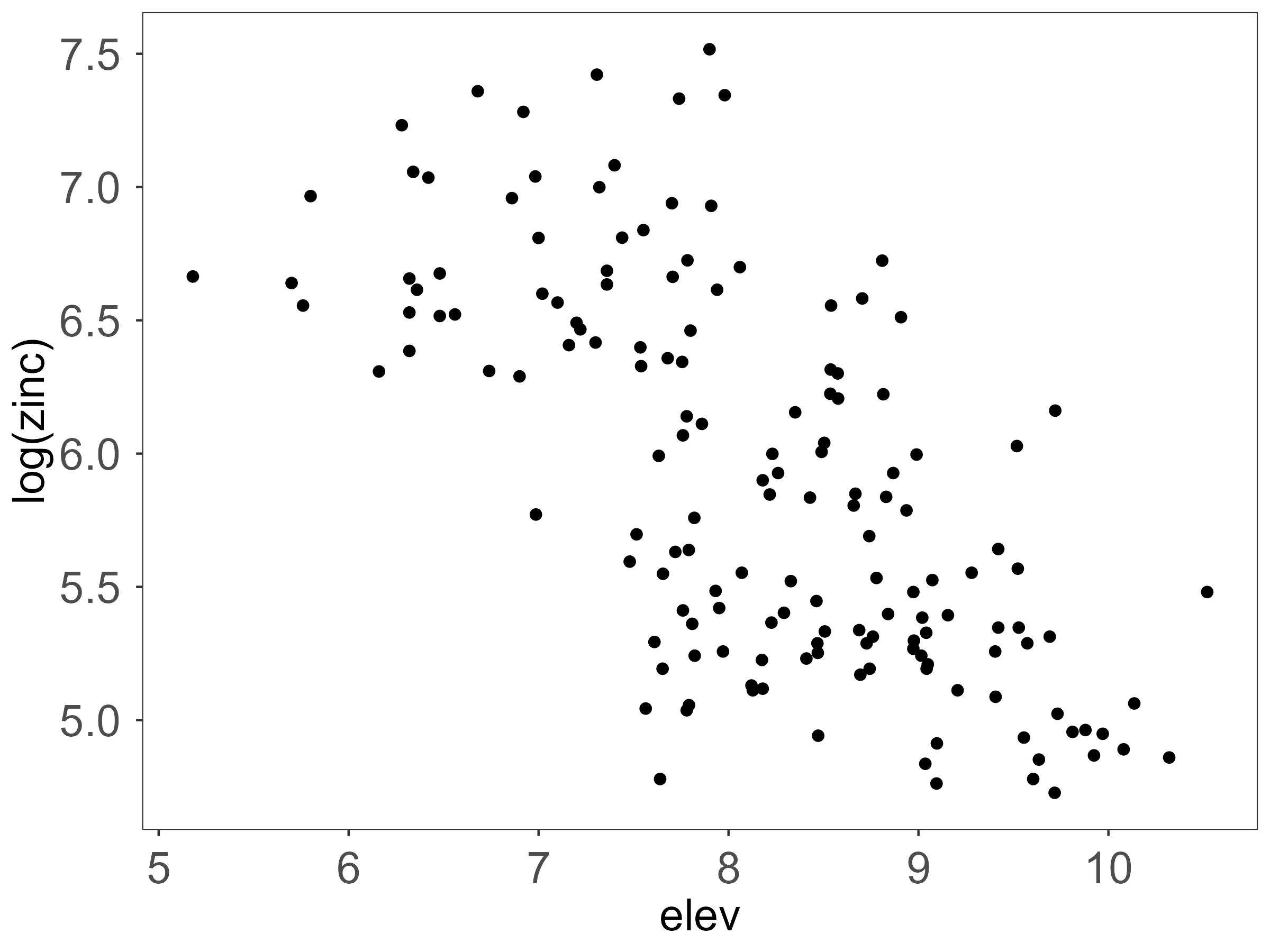}}
\caption{(a) Observed log-zinc values over the sampling locations ($w_1, w_2$); (b) Scatterplot of observed log-zinc with the covariate distance; (c) Scatterplot of observed log-zinc with the covariate elevation.}
\label{fig:obsMeuse}
\end{figure}

\begin{figure}[H]
\centering{\includegraphics[width=.47\linewidth]{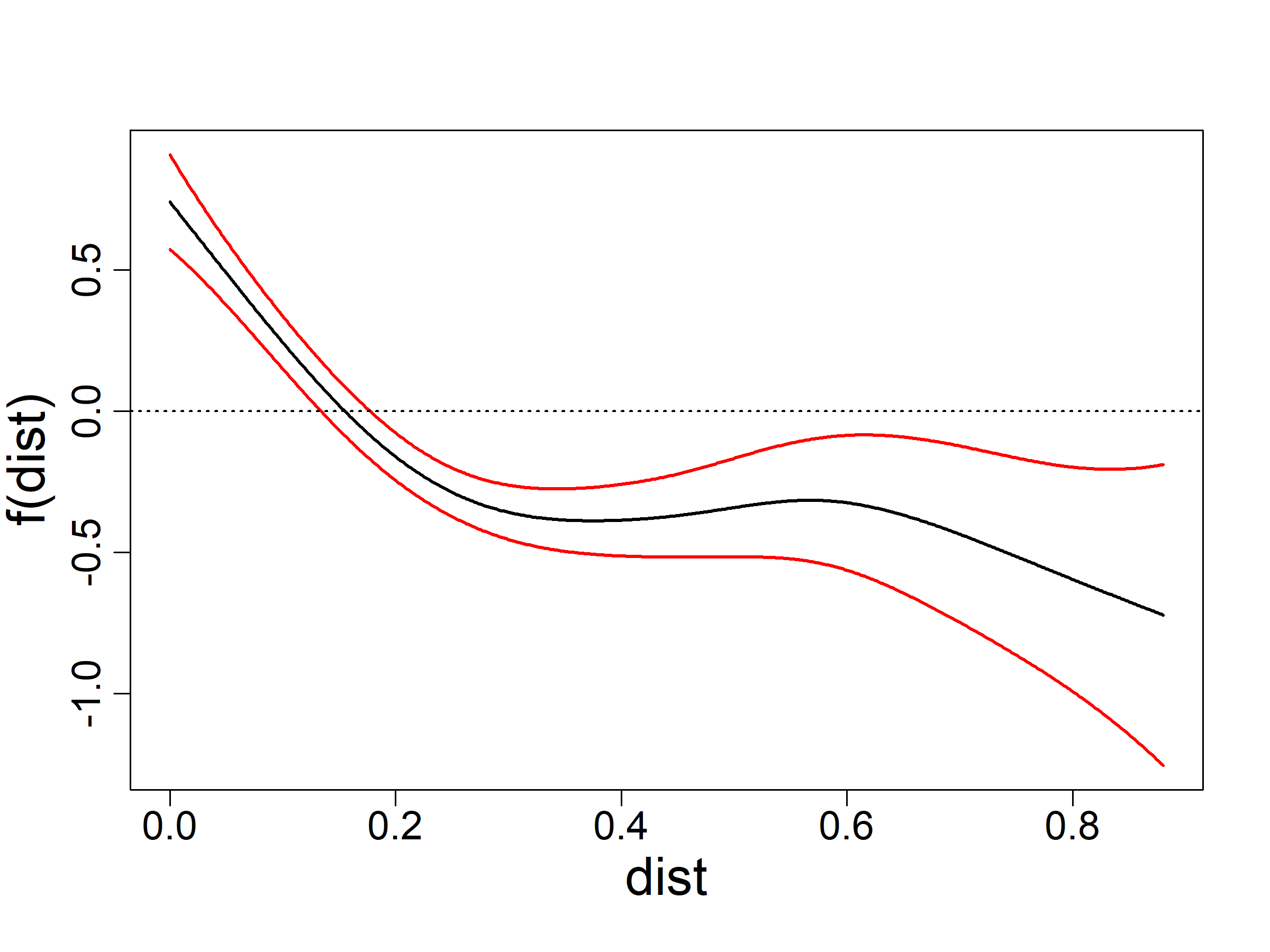}}{\includegraphics[width=.47\linewidth]{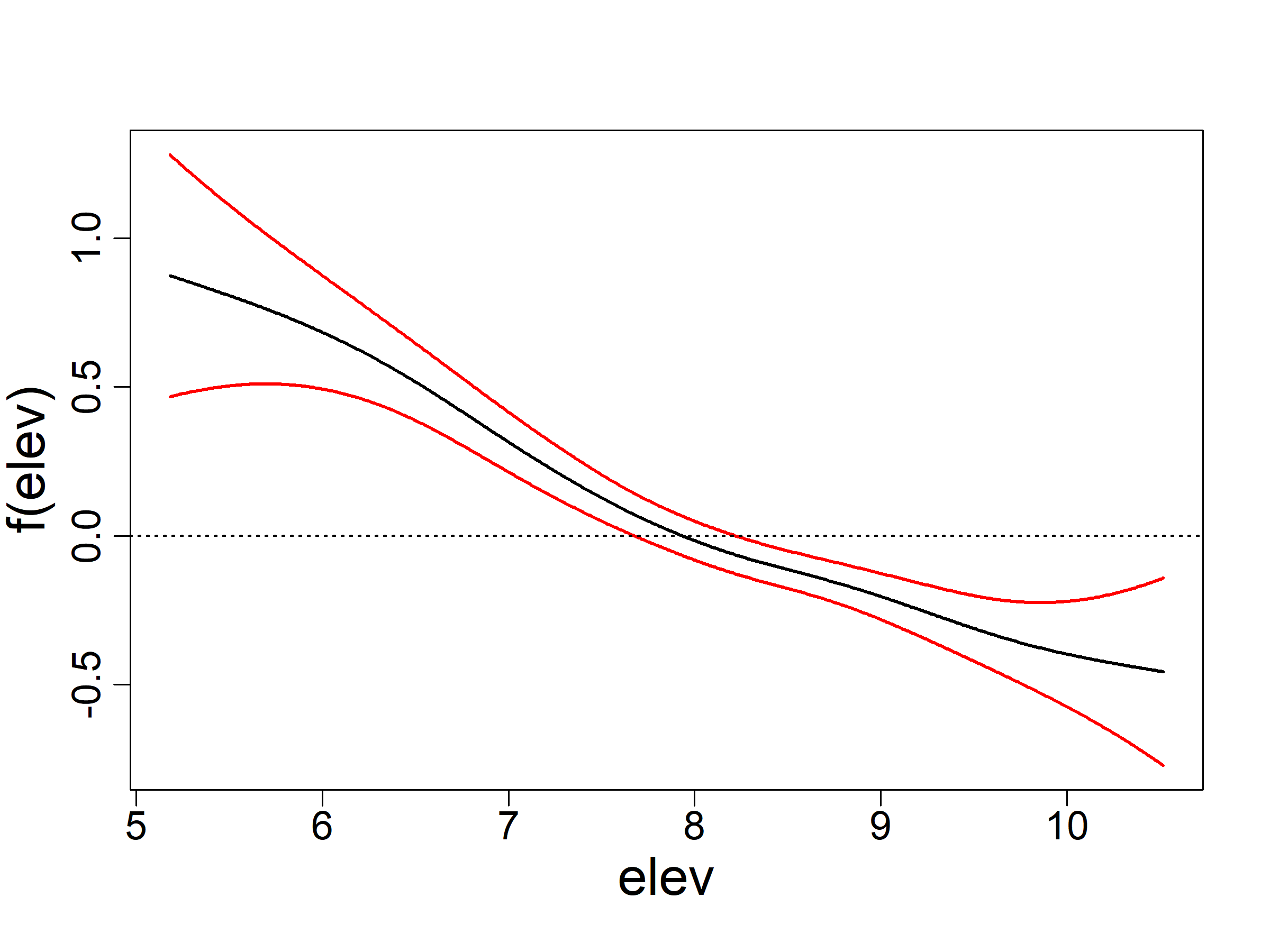}}
\caption{Estimated effects of smooth covariates on log-zinc.}
\label{fig:smooth_meuse}
\end{figure}

\begin{figure}[H]
\centering
\subfloat[]{\includegraphics[width=.47\linewidth]{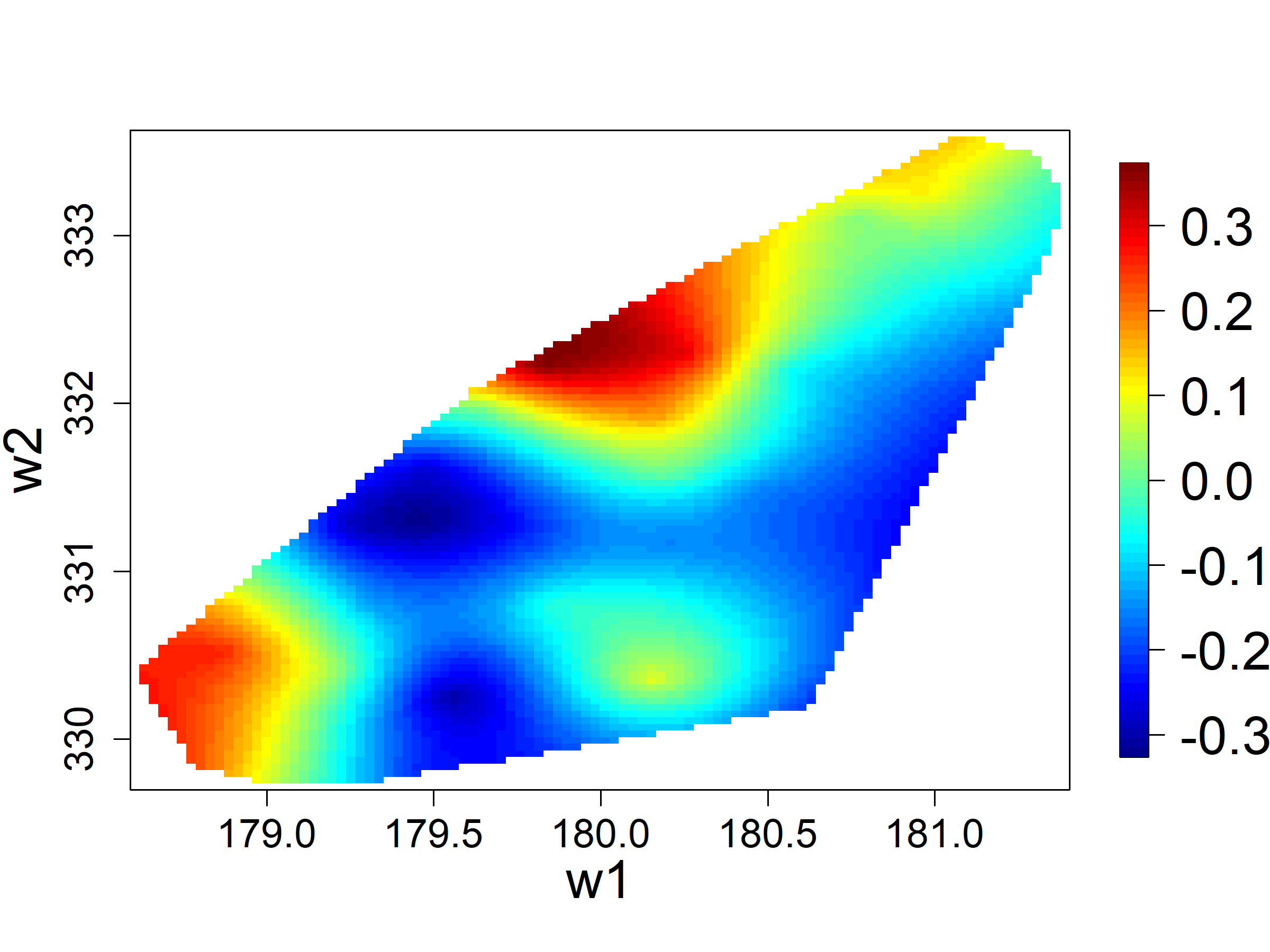}}
\hspace{0.05\linewidth}
\subfloat[]{\includegraphics[width=.47\linewidth]{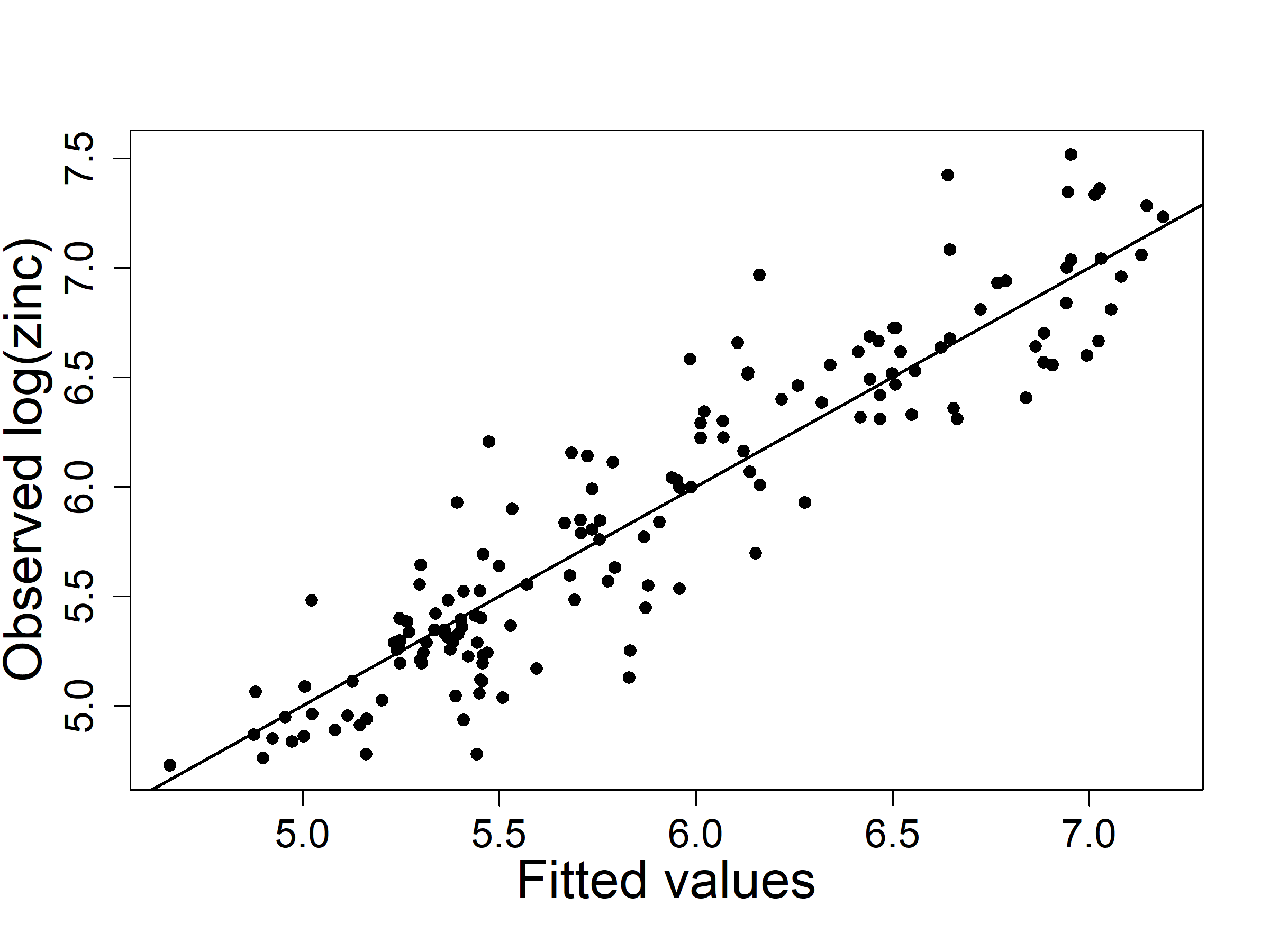}}
\caption{Results for the Meuse data using model \eqref{eqn:meusefinalmod}. (a) Estimated continuous surface for the spatial term $s(w_1,w_2)$; (b) Estimated mean vs observed log-zinc values.}
\label{fig:fitted_meuse}
\end{figure}

Figure \ref{fig:obsMeuse} displays the observed logarithm of zinc (log-zinc) values at sampling locations $(w_1, w_2)$ and scatter plots illustrating the relationship between log-zinc and the variables, distance, and elevation. The data indicates nonlinearity between distance and log-zinc values, whereas elevation suggests linearity. Classical geostatistical methods typically accommodate only the linear effects of covariates. Therefore, transformation is commonly applied (e.g. a square root transformation
 of distance) in order to fit a linear geostatistical model. One benefit of our proposed method is the direct incorporation of nonlinear covariates without the need for any transformation. The distance and elevation are included as smooth covariates, using a Gaussian model for the log-zinc, with various covariance structures. Table \ref{tab:MeuseBIC} shows that the BIC values are similar for different model covariances, with the Circular covariance having the lowest BIC.  Consequently, the model using circular covariance is examined further. Both tests for the significance of the two smooth covariates yield a $\text{p-value} < 0.0001$, indicating a statistically significant relationship between the covariates and log-zinc. Since  Figure \ref{fig:obsMeusec} indicates a linear effect of elevation, a model is fitted with elevation as a linear covariate, resulting in a BIC of -126.89, higher than the BIC (-128.94) obtained when elevation is included as a smooth covariate. Therefore, the final model includes both distance and elevation as smooth covariates given by:
\begin{equation}
\label{eqn:meusefinalmod}
    \log(zinc_i) = \beta_0 + f(dist_i) +  f(elev_i) + s(w_{1i}, w_{2i}) + \epsilon_i,
\end{equation}

for $i=1,\dots, 155$. Figure \ref{fig:smooth_meuse}  presents the estimated effects of the covariates. Figure \ref{fig:fitted_meuse} shows the estimated spatial surface and the comparison between fitted and observed log-zinc values based on model \eqref{eqn:meusefinalmod}.

\begin{table}[H]
\centering
\captionsetup{width=0.7\textwidth}
\caption{Results for Gaussian model fitted on Meuse river data.}
\label{tab:MeuseBIC}
\begin{tabular}{|l|c|}
\hline
\textbf{Covariance} & \textbf{BIC} \\ \hline
Circular            & -128.94      \\ \hline
Exponential         & -128.12      \\ \hline
Mat\'ern              & -126.61      \\ \hline
Spherical           & -127.14      \\ \hline
\end{tabular}
\end{table}

\subsection{COVID-19 vulnerability data}

The proposed negative binomial model is applied to the analysis of the COVID-19 data from Flanders and Brussels regions in Belgium from September 1, 2020, to December 31, 2020. The study area is divided into 9627 statistical sectors, each with a population ranging from a minimum of 7 to a maximum of 6082, with an average population of approximately 740 inhabitants. The population per 100 inhabitants of each statistical sector is used as an offset term in the model. The centroid of each statistical sector serves as the coordinate for the spatial analysis. The dataset includes the number of positive cases and various risk factors for each statistical sector. Variables identified in the literature as risk factors for vulnerability to COVID-19 are considered. Correlations between these variables are then examined, with only one variable retained from each pair that had an absolute Pearson correlation greater than 0.5. The final factors include median net income (\textit{med}\_\textit{inc}) \citep{rozenfeld2020, wachtler2020}, the proportion of retired people (\textit{pensinr}) \citep{rozenfeld2020, pijls2021}, the proportion of non-Belgian residents (\textit{nonBel}) \citep{hayward2021}, the proportion of single parents \citep{sung2021} (\textit{snglprn}), the yearly average black carbon level (\textit{bc}) \citep{rozenfeld2020, wu2020airpollution}, and the proportion of females (\textit{female}) \citep{wu2022gender}. All these variables are standardized.\\

Initially, all factors are considered as smooth covariates. Various covariance functions are fitted with the spherical covariance having the lowest BIC, as shown in Table \ref{tab:CovidBIC}. Therefore, further data analysis is conducted using spherical covariance, showing that all factors have statistically significant effects ($\text{p-value} < 0.0001$). The plot illustrating the estimated smooth effects is shown in Figure C.2 in the Appendix. To investigate the linear effects, each smooth covariate is subsequently replaced as a linear covariate, and the BIC is compared to that of the full model ($\text{BIC} = 1221406$), where all factors are smooth covariates. As shown in Table \ref{tab:Covid_linearBIC}, the factors \textit{med\_inc} and \textit{bc} resulted in lower BIC when included as linear covariates rather than smooth covariates.

\begin{table}[H]
\centering
\captionsetup{width=0.6\textwidth}
\caption{BIC results for the negative binomial model fitted on COVID-19 data using different covariance functions.}
\label{tab:CovidBIC}
\begin{tabular}{|l|c|}
\hline
\textbf{Covariance} & \textbf{BIC} \\ \hline
Circular            & -1221386     \\ \hline
Exponential         & -1221397     \\ \hline
Mat\'ern            & -1221187     \\ \hline
Spherical           & -1221406     \\ \hline
\end{tabular}
\end{table}

\begin{table}[H]
\centering
\captionsetup{width=0.7\textwidth}
\caption{BIC comparison for the COVID-19 data using spherical covariance when the variables are added as linear covariates.}
\label{tab:Covid_linearBIC}
\begin{tabular}{|l|c|c|}
\hline
\textbf{Variables} & \multicolumn{1}{|c|}{\textbf{BIC with linear}} & \multicolumn{1}{|c|}{\textbf{Difference from BIC}} \\
                   & \multicolumn{1}{|c|}{\textbf{covariate}}      & \multicolumn{1}{|c|}{\textbf{with smooth covariate}} \\ \hline
med\_inc           & -1221420     & -14                    \\ \hline
pensinr            & -1221388     & 18                     \\ \hline
nonBel             & -1221403     & 3                      \\ \hline
snglprn            & -1221351     & 55                     \\ \hline
bc                 & -1221422     & -16                    \\ \hline
female             & -1221388     & 18                     \\ \hline
\end{tabular}
\end{table}

\begin{figure}[H]
\centering{\includegraphics[width=.4\linewidth]{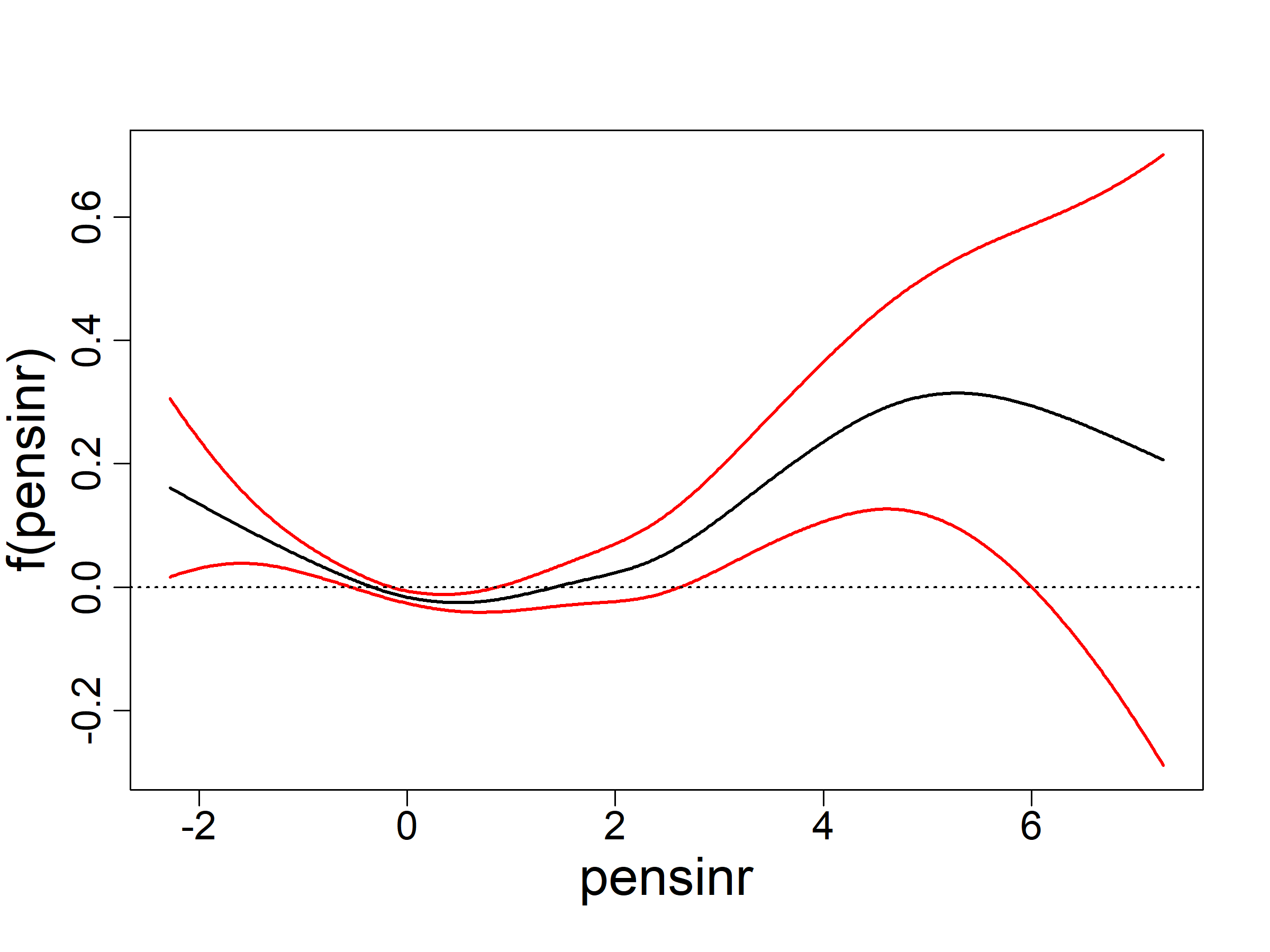}}{\includegraphics[width=.4\linewidth]{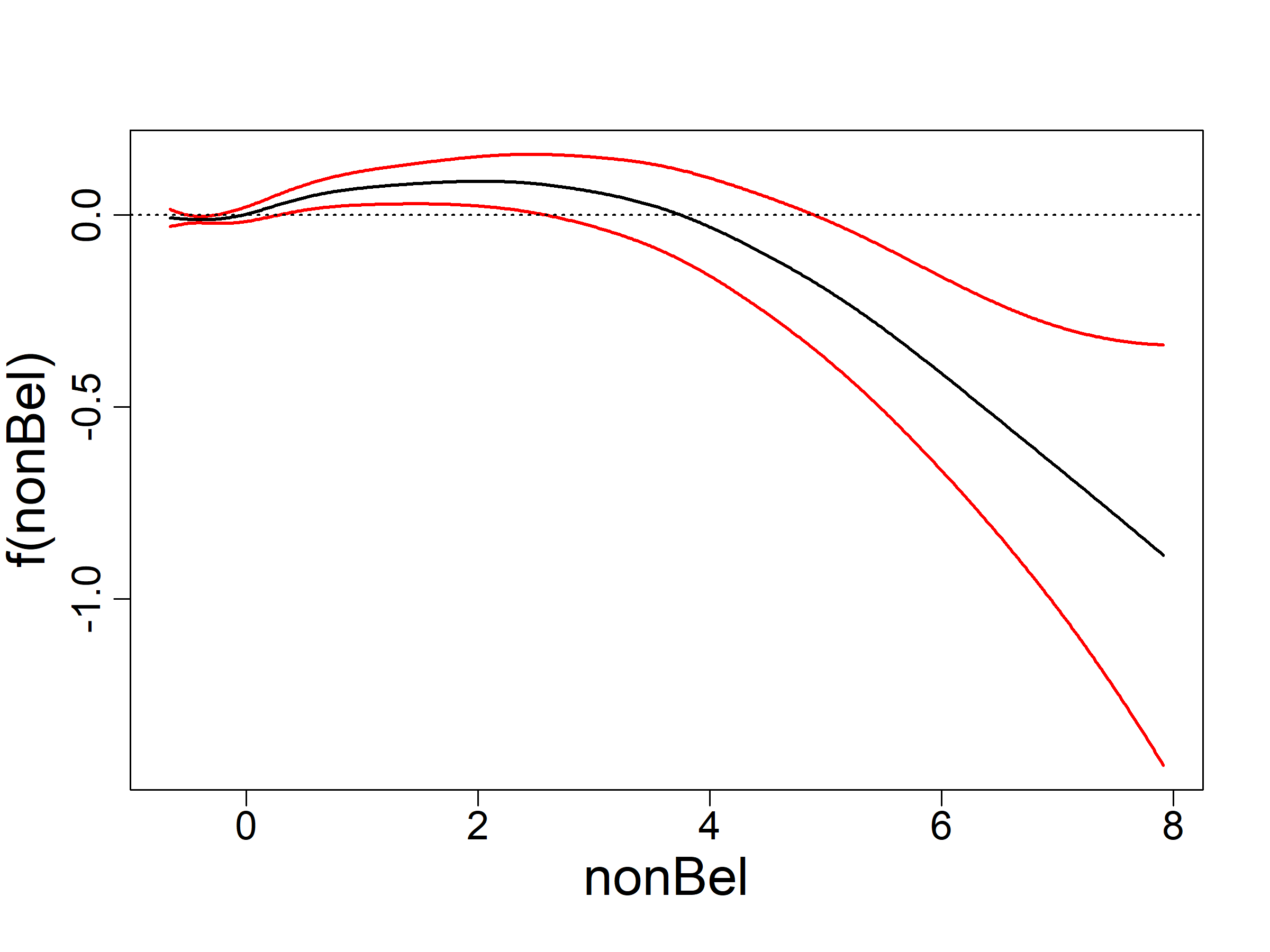}}\\
\centering{\includegraphics[width=.4\linewidth]{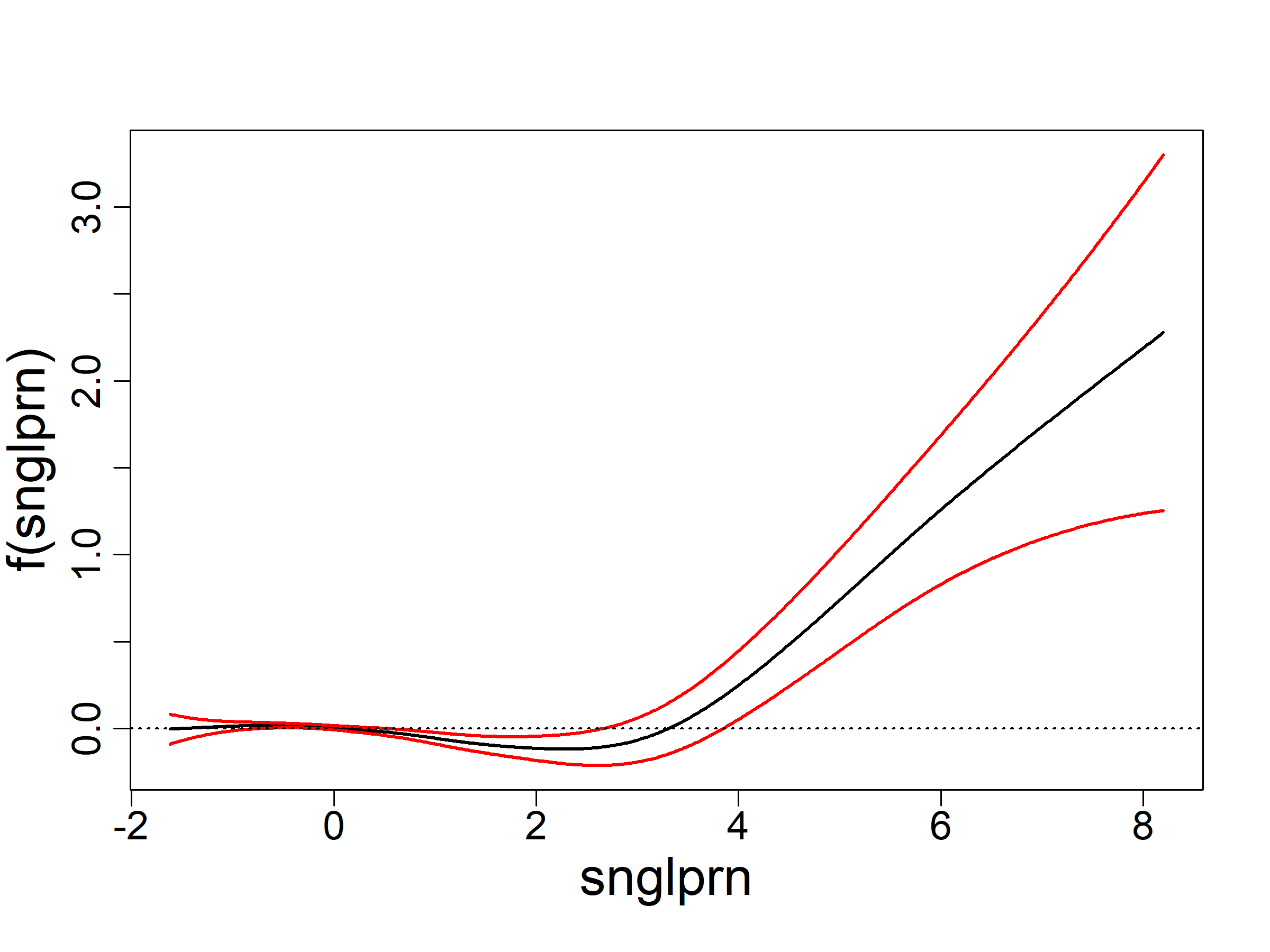}}{\includegraphics[width=.4\linewidth]{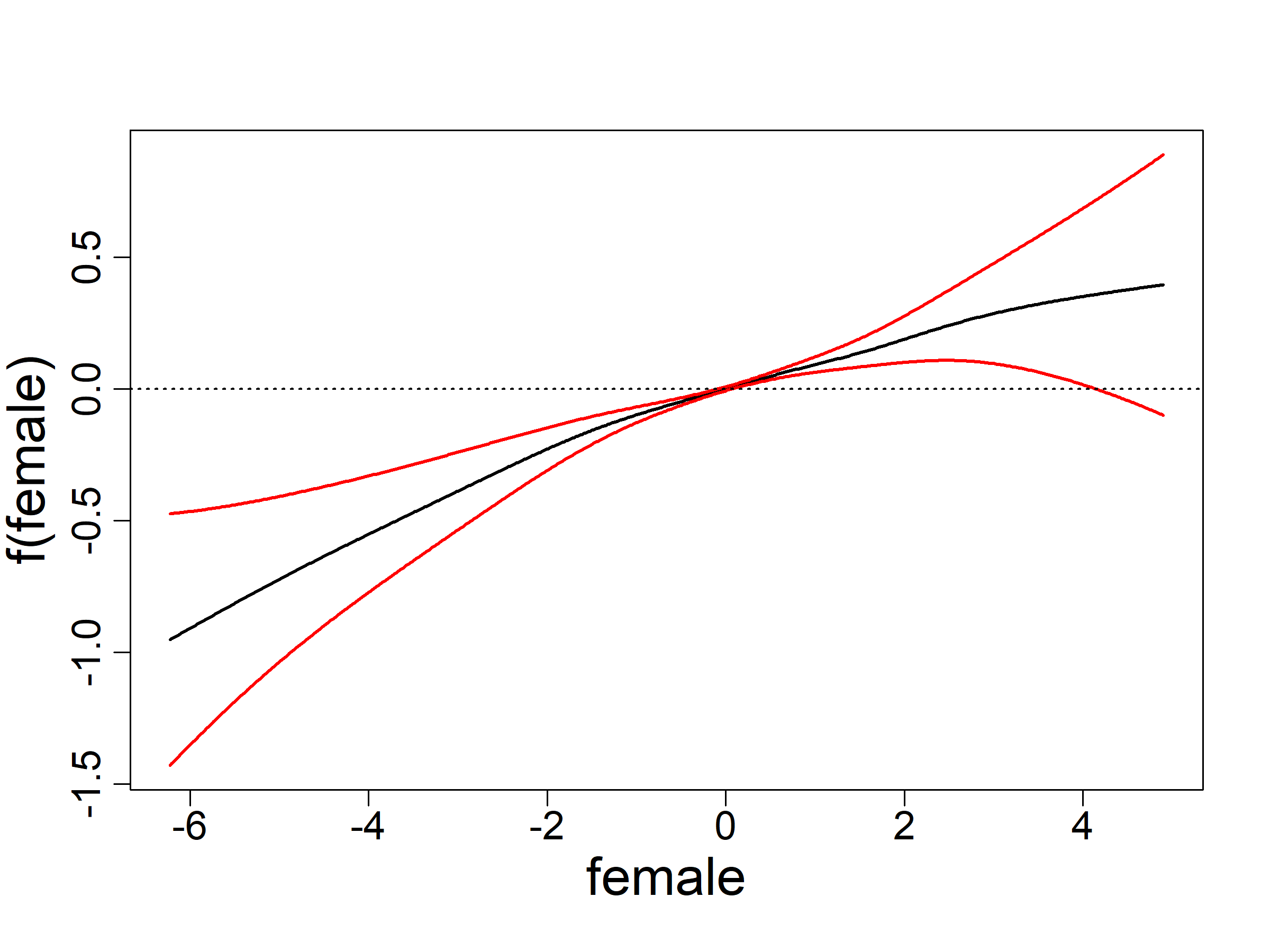}}
\caption{Estimated effects of smooth covariates for COVID-19 data using spherical covariance.}
\label{fig:smooth_COVID}
\end{figure}

\begin{figure}[H]
\centering
\subfloat[]{\includegraphics[width=.6\linewidth]{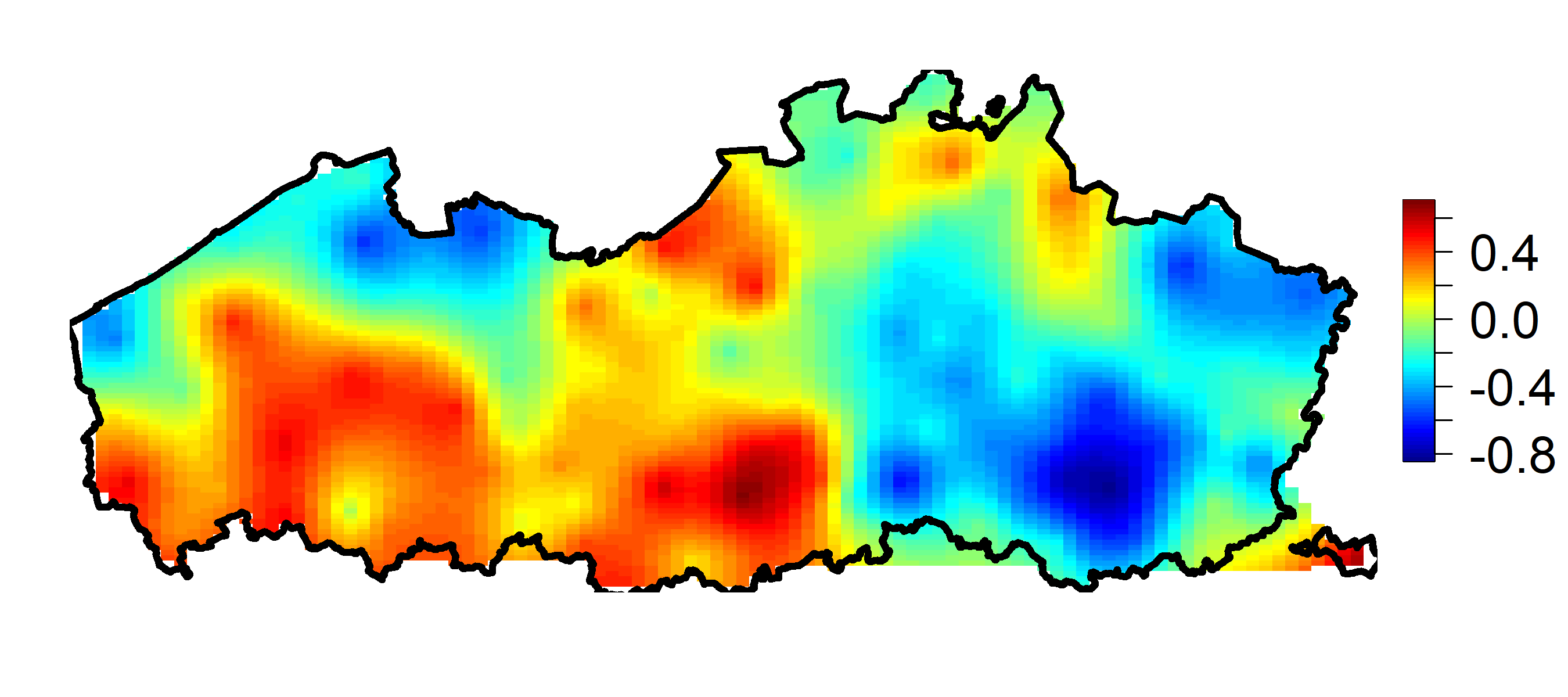 }}
\hspace{0.05\linewidth}
\subfloat[]{\includegraphics[width=.5\linewidth]{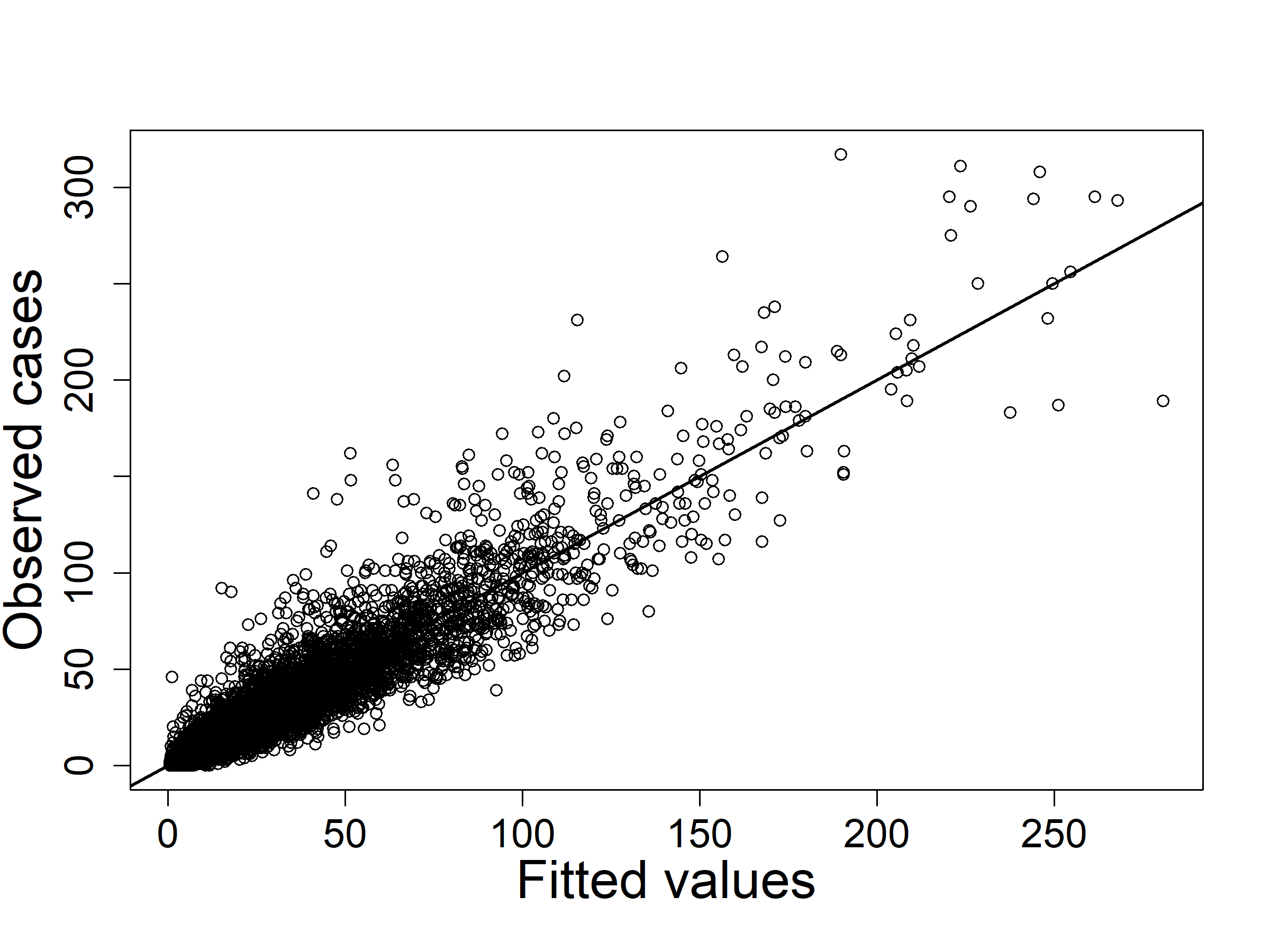 }}
\caption{Results for the COVID-19 data using model \eqref{eqn:finalmod_COVID}. (a) Estimated continuous surface for the spatial term $s(w_1,w_2)$ overlaying the map of the study region; (b) Estimated mean vs observed number of COVID-19 positive cases.}
\label{fig:fitted_covid}
\end{figure}

Therefore, the final model is given by:
\begin{align}
    \log(\mu_i) = & \beta_0 + \beta_1 med\_inc_{i} + \beta_2 bc_i + f(pensinr_i) + f(nonBel_i) + \nonumber
    \\ & f(sngplrn_i) + f(female_i) + s(w_{1i}, w_{2i}) + \log(N_i),
    \label{eqn:finalmod_COVID}
\end{align}
where $N_i$ is the offset term per 100 population for $i= 1,\dots,9627$. The estimated coefficients for the linear covariates in Table \ref{tab:lin_final_covid} indicate that areas with higher median net income are negatively correlated with the number of COVID-19 cases, while areas with high levels of black carbon are positively correlated with the number of cases. Table \ref{tab:smooth_final_covid} shows the results for the nonlinear covariates in model \eqref{eqn:finalmod_COVID} and the plot of the estimated smooth effects is shown in Figure \ref{fig:smooth_COVID}. The latter figure shows that areas with higher proportions of pensioners and single parents exhibit higher vulnerability to COVID-19 compared to the average. However, areas with average or lower proportions of these groups correspond to average vulnerability. For the covariate \textit{pensinr}, the smooth trend for higher values shows a wider credible interval, indicating higher uncertainty in the trend direction, which could potentially go up or down. This is due to the limited number of observations (only five) for \textit{pensinr} values greater than 5. Other covariates also show higher uncertainty at the right tail because of the few observations, which are right-skewed (see Figure C.1 in the Appendix). Areas with higher proportions of non-Belgians and lower proportions of females appear to be less vulnerable to COVID-19. Finally, the plot for the estimated spatial surface and the comparison between fitted and observed cases is shown in Figure \ref{fig:fitted_covid}.

\begin{table}[H]
\centering
\captionsetup{width=0.8\textwidth}
\caption{Results for the linear covariates in model \ref{eqn:finalmod_COVID}. Estimate - estimated coefficient; SE - standard error; CI lower - 95\% lower credible interval; CI upper - 95\% upper credible interval.}
\begin{tabular}{|l|c|c|c|c|}
\hline
\textbf{Variables} & \textbf{Estimate} & \textbf{SE} & \textbf{CI lower} & \textbf{CI upper} \\ \hline
(Intercept)         & 1.180             & 0.005       & 1.170              & 1.190              \\ \hline
med\_inc             & -0.073            & 0.007       & -0.088             & -0.059             \\ \hline
bc                  & 0.047             & 0.010       & 0.028              & 0.066              \\ \hline
\end{tabular}
\label{tab:lin_final_covid}
\end{table}

\begin{table}[H]
\centering
\caption{Results for the smooth covariates in model \ref{eqn:finalmod_COVID}.}
\begin{tabular}{|l|c|c|c|}
\hline
\textbf{Variables} & \textbf{ED} & $\boldsymbol{T}_r$ & \textbf{p-value} \\ \hline
pensinr             & 5.10         & 37.95       & $<0.0001$          \\ \hline
nonBel              & 4.92         & 26.27       & $<0.0001$          \\ \hline
snglprn             & 5.87         & 45.68       & $<0.0001$          \\ \hline
female              & 3.43         & 63.86       & $<0.0001$          \\ \hline
\end{tabular}
\label{tab:smooth_final_covid}
\end{table}

\section{Conclusion}
This paper presents a novel Bayesian method for geostatistical modeling that combines Laplace approximations, P-splines, and low-rank representations for spatial processes. The proposed model offers several advantages, including the ability to perform spatial interpolation and investigate the effects of both linear and nonlinear covariates. Its computational efficiency is substantially improved by using Laplace approximations and a low-rank representation of the spatial process, leading to faster computation times. Additionally, the model extends beyond Gaussian responses to accommodate Poisson and negative binomial models, thereby providing robust options for handling count data. Thus, the proposed approach not only enhances the flexibility and speed of geostatistical analyses but also broadens their applicability across different types of data. \\

Simulation studies demonstrate that the proposed model exhibits low relative bias and has credible intervals close to the nominal coverage with respect to the underlying target smooth function and spatial process. These results are consistent across all covariance functions, except for the Matérn covariance, which did not perform well in terms of credible interval coverage for the spatial component in non-symmetric two-dimensional functions. Additionally, the proposed model demonstrates good predictive interval coverage across all covariance functions, highlighting its robust predictive ability. In the presence of overdispersion, the Poisson model, as expected, results in undercoverage, but the availability of the negative binomial model within our proposed framework effectively addresses this issue.\\

The proposed model has been applied to the Meuse river data and COVID-19 vulnerability data in Belgium, demonstrating its practical applicability. The analysis of COVID-19 data reveals that areas with lower median incomes are more vulnerable to the virus, highlighting the economic disparities exacerbated by the pandemic \citep{rozenfeld2020, wachtler2020}. Additionally, areas with a higher proportion of pensioners who are at increased risk for severe COVID-19 outcomes show higher vulnerability \citep{rozenfeld2020, pijls2021}. Environmental factors, such as higher levels of black carbon, further contribute to increased susceptibility, underscoring how air pollution contributes to health disparities \citep{rozenfeld2020, wu2020airpollution}. Social demographics also play a significant role, for instance, areas with higher proportions of females \citep{wu2022gender} and single parents \citep{sung2021} are more vulnerable to the virus. Interestingly, while studies indicate that migrants are generally at high risk for COVID-19 \citep{hayward2021}, our findings suggest that areas with a high proportion of non-Belgians are less vulnerable to the disease.\\

Finally, an interesting extension of our proposed model is to assume a binomial distribution for the count response, considering the population size. This could enhance model performance in scenarios such as disease prevalence modeling, where count data naturally follows a binomial distribution due to the presence of binary outcomes (e.g. disease vs. no disease) within a given population size.

\section*{Competing Interest Statement}
The authors have declared no competing interest.

\newpage


\newpage
\section*{Appendix}

\subsection*{A. Derivations for the Gaussian case}
Assuming a Gaussian distribution on the observations $y_i(\boldsymbol{w}_i)$, we have the following geoadditive model:
\begin{equation*}
\label{eqn:gauss}
    y_i(\boldsymbol{w}_i) = \beta_0 + \beta_1x_{i1} + \cdots \beta_1x_{ip} +  f_1(s_{i1}) +  \cdots f_q(s_{iq}) + s(\boldsymbol{w}_i) + \epsilon_i, \; \epsilon_i \sim \mathcal{N}_{1}(0,\tau_{\epsilon}^{-1}).
\end{equation*}

The full Bayesian model for the Gaussian case is given by:
\begin{align*}
    & (\boldsymbol{y}|\boldsymbol{\xi}) \sim \mathcal{N}_{n}(C_{\rho}\boldsymbol{\xi}, \tau_{\epsilon}^{-1}I)\\
    & (\boldsymbol{\xi}|\boldsymbol{\lambda}, \tau_{\epsilon}, \rho)\sim \mathcal{N}_{\text{dim}(\boldsymbol{\xi})}(\boldsymbol{0},(\tau_{\epsilon}Q_{\boldsymbol{\xi}}^{\boldsymbol{\lambda}})^{-1}),\\
    & (\lambda_j|\delta_j)\sim \mathcal{G}\left(\frac{\nu}{2},\frac{\nu \delta_j}{2}\right), \quad j=1,\dots, q + 1,\\
    & \delta_j \sim \mathcal{G}(a_\delta,b_\delta), \quad j=1,\dots, q + 1,\\
    & p(\tau_{\epsilon}) \propto \frac{1}{\tau_{\epsilon}},\\
    & p(\rho) \propto \frac{1}{\rho}.
\end{align*}

The conditional posterior of $\boldsymbol{\xi}$ can be written as:
\begin{align*}
    p(\boldsymbol{\xi}| \boldsymbol{\lambda},\tau_{\epsilon}, \rho,\mathcal{D}) \propto & \; \mathcal{L}(\boldsymbol{\xi}, \tau_{\epsilon}, \rho | \mathcal{D}) \times  p(\boldsymbol{\xi}| \boldsymbol{\lambda},\tau_{\epsilon}, \rho)\\
    \propto & \exp(-0.5\tau_{\epsilon}||\boldsymbol{y}-C_{\rho}\boldsymbol{\xi}||^{2}) \times \exp(-0.5\tau_{\epsilon}\boldsymbol{\xi}^{\top}Q_{\boldsymbol{\xi}}^{\boldsymbol{\lambda}}\boldsymbol{\xi}).
\end{align*}
It can be shown that $p(\boldsymbol{\xi}| \boldsymbol{\lambda},\tau_{\epsilon}, \rho,\mathcal{D}) \propto \exp\left(0.5\boldsymbol{\xi} \widehat{\Sigma}_{\boldsymbol{\lambda}}^{-1}\boldsymbol{\xi} -2\boldsymbol{\xi} \widehat{\Sigma}_{\boldsymbol{\lambda}}^{-1}\widehat{\boldsymbol{\xi}}_{\boldsymbol{\lambda}}\right)$,
where $\widehat{\boldsymbol{\xi}}_{\boldsymbol{\lambda}} = (C_{\rho}^{\top}C_{\rho} + Q_{\boldsymbol{\xi}}^{\boldsymbol{\lambda}})^{-1} C_{\rho}^{\top}\boldsymbol{y}$ and $\widehat{\Sigma}_{\boldsymbol{\lambda}} = \tau_{\epsilon}^{-1} (C_{\rho}^{\top}C_{\rho} + Q_{\boldsymbol{\xi}}^{\boldsymbol{\lambda}})^{-1}$. This can be recognized as proportional to a multivariate Gaussian distribution with mean vector $\widehat{\boldsymbol{\xi}}_{\boldsymbol{\lambda}}$ and covariance matrix $\widehat{\Sigma}_{\boldsymbol{\lambda}}$. Therefore, the conditional posterior of $\boldsymbol{\xi}$ is a Gaussian density given by
$(\boldsymbol{\xi}| \boldsymbol{\lambda},\tau_{\epsilon}, \rho,\mathcal{D}) \sim \mathcal{N}_{\text{dim}(\boldsymbol{\xi})}(\widehat{\boldsymbol{\xi}}_{\boldsymbol{\lambda}}, \widehat{\Sigma}_{\boldsymbol{\lambda}}).$ \\

\newpage

\textbf{\large Posterior distribution of hyperparameters}\\

Using Bayes' theorem, the marginal joint posterior of hyperparemeters $\boldsymbol{\lambda}$, $\tau_{\epsilon}$ and $\rho $ is given by:
\begin{align*}
     p(\boldsymbol{\lambda}, \tau_{\epsilon}, \rho|\mathcal{D})
     & \propto \frac{\mathcal{L}(\boldsymbol{\xi}, \tau_{\epsilon}, \rho ;\mathcal{D})p(\boldsymbol{\xi}|\boldsymbol{\lambda}, \tau_{\epsilon}, \rho)p(\boldsymbol{\lambda}|\boldsymbol{\delta}) p(\tau_{\epsilon}) p(\rho) }{p(\boldsymbol{\xi}|\boldsymbol{\lambda}, \tau_{\epsilon}, \rho,\mathcal{D})}.
\end{align*}

The above posterior can be approximated by evaluating $\boldsymbol{\xi}$ at the mode $\widehat{\boldsymbol{\xi}}_{\boldsymbol{\lambda}}$. Note also that $|Q_{\boldsymbol{\xi}}^{\boldsymbol{\lambda}}| = |V_{\beta}| \times |\lambda_1 P| \times \cdots \times |\lambda_q P| \times |\lambda_{q+1} \Omega_{\rho}| \propto \lambda_1^{K} \times \cdots \times \lambda_q^{K} \times \lambda_{q+1}^{S}|\Omega_{\rho}|$. Hence, the approximated marginal posterior of the hyperparameter is given by
\begin{align*}
     \widetilde{p}(\boldsymbol{\lambda}, \tau_{\epsilon}, \rho|\mathcal{D})
     & \propto  \tau_{\epsilon}^{\frac{n}{2}} \exp\left(-0.5\tau_{\epsilon}||\boldsymbol{y} - C_{\rho}\widehat{\boldsymbol{\xi}}_{\boldsymbol{\lambda}}||^2 \right) \times \tau_{\epsilon}^{\frac{\text{\tiny dim}(\widehat{\boldsymbol{\xi}}_{\boldsymbol{\lambda}})}{2}} |Q_{\boldsymbol{\xi}}^{\boldsymbol{\lambda}}|^{\frac{1}{2}}\exp(-0.5\tau_{\epsilon}\widehat{\boldsymbol{\xi}}_{\boldsymbol{\lambda}}^{\top}Q_{\boldsymbol{\xi}}^{\boldsymbol{\lambda}}\widehat{\boldsymbol{\xi}}_{\boldsymbol{\lambda}}) \nonumber \\
     & \times \prod_{j=1}^{q+1}(\delta_j)^{\frac{\nu}{2}} (\lambda_j)^{\frac{\nu}{2}-1} \exp\left(-\frac{\nu \delta_j}{2}\lambda_j\right) \times \prod_{j=1}^{q+1}(\delta_j)^{a_{\delta}-1} \exp(-b_{\delta}\delta_j) \nonumber\\
     & \times \frac{1}{\tau_{\epsilon}} \times \frac{1}{\rho} \times |\widehat{\Sigma}_{\boldsymbol{\lambda}}|^{\frac{1}{2}} \times \tau_{\epsilon}^{-\frac{\text{\tiny dim}(\widehat{\boldsymbol{\xi}}_{\boldsymbol{\lambda}})}{2}} |C_{\rho}^{\top}C_{\rho} + Q_{\boldsymbol{\xi}}^{\boldsymbol{\lambda}}|^{-\frac{1}{2}} \nonumber\\
     & = \tau_{\epsilon}^{\frac{n}{2} - 1} \exp\left( -0.5\tau_{\epsilon} (||\boldsymbol{y} - C_{\rho}\widehat{\boldsymbol{\xi}}_{\boldsymbol{\lambda}}||^2 + \widehat{\boldsymbol{\xi}}_{\boldsymbol{\lambda}}^{\top}Q_{\boldsymbol{\xi}}^{\boldsymbol{\lambda}}\widehat{\boldsymbol{\xi}}_{\boldsymbol{\lambda}}) \right) \times \prod_{j=1}^{q}(\lambda_j)^{\frac{K + \nu}{2} - 1} (\lambda_{q+1})^{\frac{S + \nu}{2} - 1} \nonumber \\
     & \times \frac{1}{\rho} |\Omega_{\rho}|^{-\frac{1}{2}} |C_{\rho}^{\top}C_{\rho} + Q_{\boldsymbol{\xi}}^{\boldsymbol{\lambda}}|^{-\frac{1}{2}} \prod_{j=1}^{q+1}(\delta_j)^{\frac{\nu}{2} + a_{\delta}-1} \exp\left(-\left(\frac{\nu \lambda_j}{2} + b_{\delta}\right)\delta_j\right). \quad \quad \quad \text{Eq. (A.1) } 
\end{align*}

The terms in the last line involving $\tau_{\epsilon}$ and $\delta_j$ can be recognized as a kernel of a gamma density. Thus, integrating $\widetilde{p}(\boldsymbol{\lambda}, \tau_{\epsilon}, \rho|\mathcal{D})$ successively with respect to $\tau_{\epsilon}$ and $\delta_j$ for $j=1,..,q+1$ yields
\begin{align*}
    \widetilde{p}(\boldsymbol{\lambda}, \rho | D) \propto & (||\boldsymbol{y}-C_{\rho}\widehat{\boldsymbol{\xi}}_{\boldsymbol{\lambda}}||^{2} + \widehat{\boldsymbol{\xi}}_{\boldsymbol{\lambda}}^{\top}Q_{\boldsymbol{\xi}}^{\boldsymbol{\lambda}}\widehat{\boldsymbol{\xi}}_{\boldsymbol{\lambda}} )^{-\frac{n}{2}} \; \times \rho^{-1}|\Omega_{\rho}|^{\frac{1}{2}} \; \times |C_{\rho}^{\top}C_{\rho} + Q|^{-\frac{1}{2}} \nonumber\\
    & \times \prod_{j=1}^{q}(\lambda_j)^{\frac{K + \nu}{2} - 1} (\lambda_{q+1})^{\frac{S + \nu}{2} - 1}  \times \prod_{j=1}^{q+1} \left(\frac{\nu \lambda_j}{2} + b_{\delta}\right)^{-\frac{\nu}{2} + a_{\delta}}. \label{eqn:jointhylambdaphi}
\end{align*}

Let $v = (v_1, \dots, v_{q+1})^{\top} = (\log(\lambda_1), \dots, \log(\lambda_{q+1})^{\top}$ and $v_{\rho} = \log(\rho)$. Using the method of transformation, the Jacobian of the transformation is given by $J = \exp(v_{\rho}) \prod_{j=1}^{q+1}\exp(v_j)$. Then the joint posterior of $v$ and $v_{\rho}$ is given by

\begin{align*}
    \widetilde{p}(\boldsymbol{v}, v_{\rho} | D) \propto & (||\boldsymbol{y}-C_{v_{\rho}}\widehat{\boldsymbol{\xi}}_{\boldsymbol{\lambda}}||^{2} + \widehat{\boldsymbol{\xi}}_{\boldsymbol{\lambda}}^{\top}Q_{\boldsymbol{\xi}}^{\boldsymbol{\lambda}}\widehat{\boldsymbol{\xi}}_{\boldsymbol{\lambda}} )^{-\frac{n}{2}} \; \times |\Omega_{v_{\rho}}|^{\frac{1}{2}} \; \times |C_{v_{\rho}}^{\top}C_{v_{\rho}} + Q_{\boldsymbol{\xi}}^{\boldsymbol{\lambda}}|^{-\frac{1}{2}}\\
    & \times \prod_{j=1}^{q}(\exp(v_j))^{\frac{K + \nu}{2}} (\exp(v_{q+1}))^{\frac{S + \nu}{2}} \times \prod_{j=1}^{q+1} \left(\frac{\nu \exp(v_j)}{2} + b_{\delta}\right)^{-\frac{\nu}{2} + a_{\delta}}.
\end{align*}

Moreover, the log-posterior of $v$ and $v_{\rho}$ is
\begin{align*}
    \log(\widetilde{p}(\boldsymbol{v}, v_{\rho} | D)) & \dot{=} -\frac{n}{2} \log(||\boldsymbol{y}-C_{v_{\rho}}\widehat{\boldsymbol{\xi}}_{\boldsymbol{\lambda}}||^{2} + \widehat{\boldsymbol{\xi}}_{\boldsymbol{\lambda}}^{\top}Q_{\boldsymbol{\xi}}^{\boldsymbol{\lambda}}\widehat{\boldsymbol{\xi}}_{\boldsymbol{\lambda}}) + \frac{1}{2} \log|\Omega_{v_{\rho}}| - \frac{1}{2} \log |C_{v_{\rho}}^{\top}C_{v_{\rho}} + Q_{\boldsymbol{\xi}}^{\boldsymbol{\lambda}}| \nonumber\\
    & \quad + \sum_{j=1}^{q}\left(\frac{K + \nu}{2} v_j\right) + \left(\frac{S + \nu}{2} v_{q+1}\right) - \sum_{j=1}^{q+1} \left(\frac{\nu}{2} + a_{\delta}\right) \log \left( \frac{\nu \exp(v_j)}{2} + b_{\delta}\right).
\end{align*}

Furthermore, from Eq. (A.1), the posterior of $\tau_{\epsilon}$ conditional on  $\boldsymbol{\lambda}$ and $\rho$ is a gamma density given by $(\tau_{\epsilon}|\boldsymbol{\lambda},\rho,\mathcal{D}) \sim \mathcal{G}\left(\frac{n}{2}, 0.5(||\boldsymbol{y}-C_{\rho}\widehat{\boldsymbol{\xi}}_{\boldsymbol{\lambda}}||^{2} + \widehat{\boldsymbol{\xi}}_{\boldsymbol{\lambda}}^{\top}Q_{\boldsymbol{\xi}}^{\boldsymbol{\lambda}}\widehat{\boldsymbol{\xi}}_{\boldsymbol{\lambda}} )\right)$.

\subsection*{B. Simulation study comparing the proposed model with the classical kriging approach}

The proposed model for the Gaussian data is compared with the classical kriging approach using the \texttt{likfit()} function in the R package \texttt{geoR}. A model with constant mean and spatial term only is assumed for this simulation scenario. Specifically, the observations are generated from the model $y = \mu + \epsilon$, where $\mu = \beta_0 + s(w_1,w_2)$, $\beta_0 = 3$, $\epsilon \sim \mathcal{N}(0, \sqrt{0.10})$. The spatial term $s(w_1,w_2)$ is simulated from a Gaussian Random Field (GRF) using the \texttt{grf()} function in R with zero mean and different forms of covariances (circular, exponential, Mat\'ern, spherical), with sill parameter $\lambda_{\text{spat}}^{-1}= 0.5$, and range parameter $\rho^{-1} = 0.15$. Results are presented in Table A.1.

\begin{table}[H]
\centering
\captionsetup{labelformat=empty ,width=0.8\textwidth}
\caption{Table B.1: Results comparing the proposed low-rank Gaussian model and the classical kriging approach with spatial component only. $\mathcal{M}_1$ - proposed method; $\mathcal{M}_2$ - using \texttt{likfit()} function in R.}
\label{tab:likfit}
\begin{tabular}{|l|cccc|cc|}
\hline
\textbf{}   & \multicolumn{4}{c|}{\textbf{$\mu$}}                                                             & \multicolumn{2}{c|}{y}    \\ \hline
            & \multicolumn{2}{c|}{Bias}                                  & \multicolumn{2}{c|}{(\%)Bias}      & \multicolumn{2}{c|}{PI Coverage}      \\ \hline
            & \multicolumn{1}{c|}{$\mathcal{M}_1$}   & \multicolumn{1}{c|}{$\mathcal{M}_2$}  & \multicolumn{1}{c|}{$\mathcal{M}_1$} & $\mathcal{M}_2$ & \multicolumn{1}{c|}{$\mathcal{M}_1$} & $\mathcal{M}_2$ \\ \hline
Circular    & \multicolumn{1}{c|}{-0.0061} & \multicolumn{1}{c|}{0.0037} & \multicolumn{1}{c|}{10.90}  & 7.43  & \multicolumn{1}{c|}{95.81} & 94.98 \\ \hline
Exponential & \multicolumn{1}{c|}{-0.0039} & \multicolumn{1}{c|}{0.0043} & \multicolumn{1}{c|}{8.58}  & 6.61  & \multicolumn{1}{c|}{95.75} & 95.14 \\ \hline
Mat\'ern      & \multicolumn{1}{c|}{-0.0023} & \multicolumn{1}{c|}{0.0011} & \multicolumn{1}{c|}{5.70}  & 2.49  & \multicolumn{1}{c|}{96.04} & 95.09 \\ \hline
Spherical   & \multicolumn{1}{c|}{-0.0067} & \multicolumn{1}{c|}{0.0039} & \multicolumn{1}{c|}{11.72}  & 7.81  & \multicolumn{1}{c|}{95.62} & 95.02 \\ \hline
\end{tabular}
\end{table}

\begin{table}[h]
\centering
\captionsetup{labelformat=empty,width=0.8\textwidth}
\caption{Table B.2: Comparison of computation time (in seconds) with 10 evaluations for Gaussian data without covariates.}
\begin{tabular}{|l|c|c|c|c|}
\hline
 & Min & Mean & Median & Max \\
\hline
Proposed method & 0.87 & 1.03 & 0.98 & 1.57 \\
\hline
Classical kriging approach (\texttt{likfit()}) & 108.19 & 109.82 & 109.48 & 112.80 \\
\hline
\end{tabular}
\end{table}

\subsection*{C. Additional results for the data applications}

\begin{figure}[H]
    \centering
    \includegraphics[width=1\textwidth]{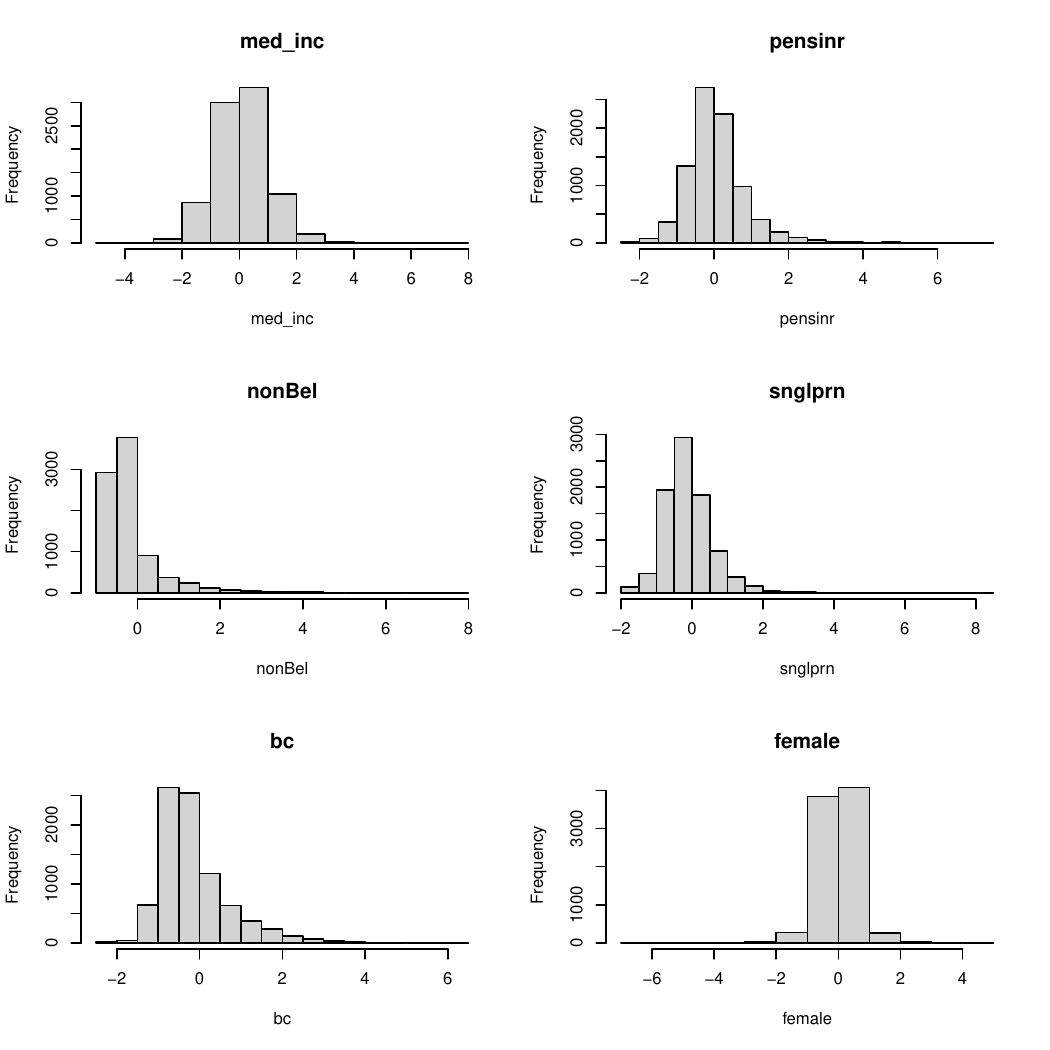}
    \caption*{Figure C.1: Histogram of the factors considered in the analysis of COVID-19 data.}
\end{figure}

\begin{figure}[H]
    \centering
    \includegraphics[width=1\textwidth]{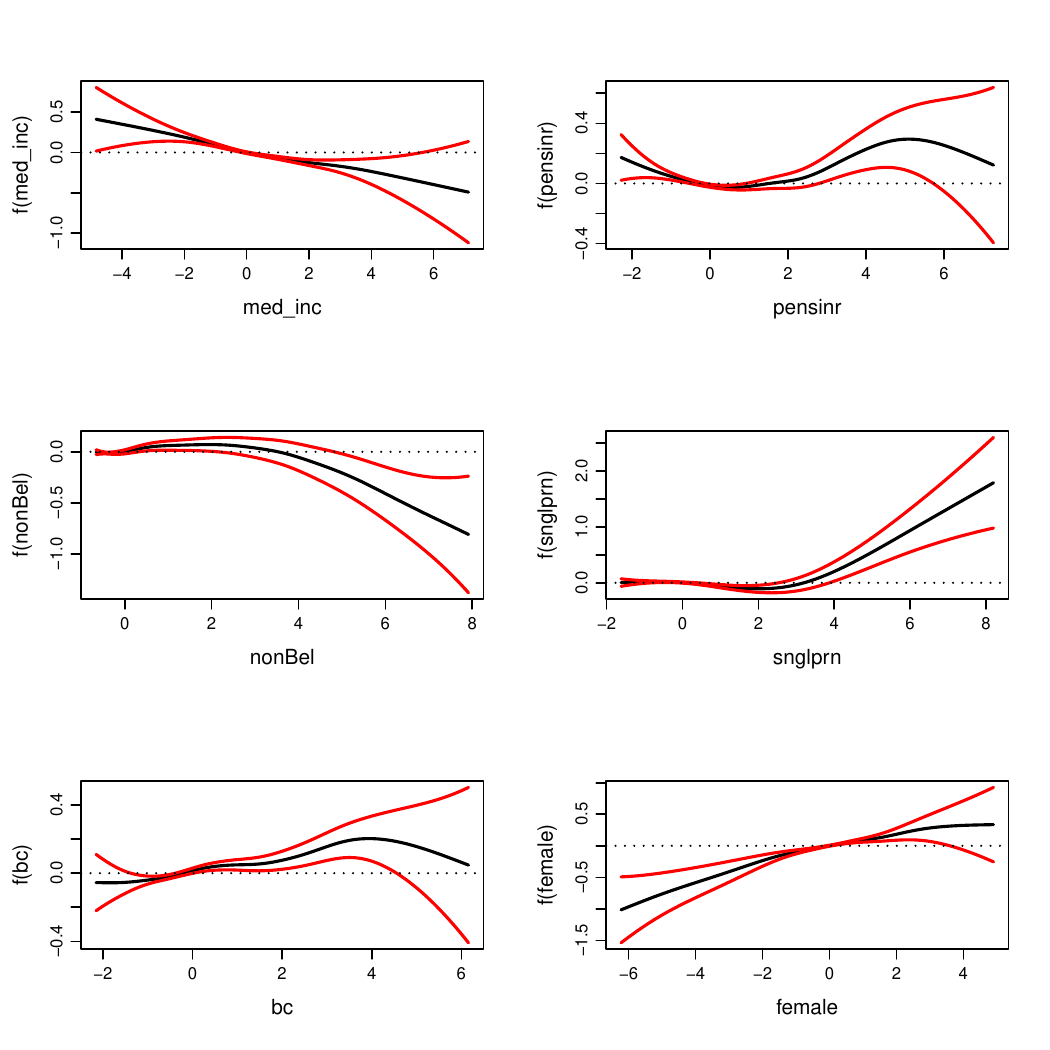}
    \caption*{Figure C.2: Estimated smooth effects for the COVID-19 data using spherical covariance.}
\end{figure}

\end{document}